\documentclass[10pt,letter,twoside]{IEEEtran}
\usepackage{amsmath}
\usepackage{booktabs}
\usepackage{amsfonts}
\usepackage{amssymb}
\usepackage{wrapfig}
\usepackage{psfrag}
\usepackage{epstopdf}
\usepackage{cite}
\usepackage{graphicx}
\usepackage{makecell}
\usepackage{subfigure}
\usepackage{threeparttable}
\usepackage[normalem]{ulem}
\usepackage{cases}
\usepackage{subeqnarray}
\usepackage{color}
\usepackage{hyperref}
\usepackage{algorithmic}
\usepackage{algorithm}

\def\QED{\hfill $\blacksquare$}
\allowdisplaybreaks
\date{}
\begin{document}
\title{Joint Secrecy Rate Achieving and Authentication Enhancement via Tag-based Encoding in Chaotic UAV Communication Environment}
\author{Junjie Wang, Fang Fang, \emph{Senior Member, IEEE}, Gangtao Han, Ning Wang, and Xianbin Wang, \emph{Fellow, IEEE}
\thanks{G. Han's work is supported in parts by National Natural Science Foundation of China under Grant 62101504 and Program of Song Shan Laboratory (221100211300-01,241110210100, included in the management of Major Science and Technology Program of Henan Province) and SongShan Laboratory Foundation (YYJC022022002). (\emph{Corresponding author: Gangtao Han.})}
\thanks{J. Wang, G. Han, and N. Wang are with the School of Electrical and Information Engineering, Zhengzhou University, China (email:jjwang@gs.zzu.edu.cn;iegthan@zzu.edu.cn;ienwang@zzu.edu.cn).}
\thanks{F. Fang and X. Wang are with the Department of Electrical and Computer Engineering, and Fang Fang is also with the Department of Computer Science, Western University, London, ON N6A 3K7, Canada.  (email: $\left\{\right.$fang.fang and xianbin.wang$\left.\right\}$@uwo.ca).}
}
\maketitle

\begin{abstract}
Secure communication is crucial in many emerging systems enabled by unmanned aerial vehicle (UAV) communication networks. 
To protect legitimate communication in a chaotic UAV environment, where both eavesdropping and jamming become straightforward from multiple adversaries with line-of-sight signal propagation, a new reliable and integrated physical layer security mechanism is proposed in this paper for a massive multiple-input-multiple-output (MIMO) UAV system.
Particularly, a physical layer fingerprint, also called a tag, is first embedded into each message for authentication purpose. 
We then propose to reuse the tag additionally as a reference to encode each message to ensure secrecy for confidentiality enhancement at a low cost. 
Specifically, we create a new dual-reference symmetric tag generation mechanism by inputting an encoding-insensitive feature of plaintext along with the key into a hash function. 
At a legitimate receiver, an expected tag, reliable for decoding, can be symmetrically regenerated based on the received ciphertext, and authentication can be performed by comparing the regenerated reference tag to the received tag. 
However, an illegitimate receiver can only receive the fuzzy tag which can not be used to decode the received message. 
Additionally, we introduce artificial noise (AN) to degrade eavesdropping to further decrease message leakage. 
To verify the efficiency of our proposed tag-based encoding (TBE) scheme, we formulate two optimization problems including ergodic sum secrecy rate maximization and authentication fail probability minimization. 
The power allocation solutions are derived by difference-of-convex (DC) programming and the Lagrange method, respectively. 
The simulation results demonstrate the superior performance of the proposed TBE approach compared to the prior AN-aided tag embedding scheme.

\end{abstract}

\section{Introduction}
Secure communication among unmanned aerial vehicles (UAVs) provides essential support for the successful execution of mission critical tasks, but the open nature of the wireless channel causes vulnerabilities to malicious threats~\cite{Tutorial1, Tutorial2, Tutorial3}. 
There are two main security issues in wireless environment, i.e., eavesdropping and jamming. 
Eavesdropping means that illegitimate receivers could intercept the transmitted messages, while jamming indicates that malicious transmitters persistently send false or useless messages to disrupt legitimate receivers~\cite{secrecy, authentication}. 
To mitigate these threats, secrecy and authentication techniques are essential for improving the security of UAV communications. 
These challenges are exacerbated in UAV networks due to the unique signal propagation environment. 

To overcome these challenges, physical layer security (PLS) aims to improve secrecy capacity and access control at the fundamental layer of communication protocol. 
By leveraging the reciprocity, randomness, and unclonability of wireless channels, illegitimate signal reception at eavesdroppers and legitimate transmitter authentication could be mitigated~\cite{PLS1, PLS2}. 
Furthermore, PLS can not only serve as the supplement and enhancement to cryptographic-based schemes in upper layers, but also it can reduce security overhead and communication latency~\cite{low_latency, supplement}. 
These advantages have attracted extensive investigations into utilizing PLS to protect UAV communications from the secrecy capability achieving and authentication enhancement perspectives~\cite{UAV-friendly jamming STAR-RIS, UAV-friendly jamming UAV NOMA, UAV_relay_beamforming, encryption, key_generation, UAV_transmitter_AN_beamforming, authentication_Xie, CSI, CSI2, CSI_CFO_PN, PLA_CSI, UAV_conference, Paul_part1, Paul_part2}.

Specifically, secrecy capability achieving techniques at the physical layer consist of artificial noise (AN), precoding (beamforming), and encryption (encoding)~\cite{UAV-friendly jamming STAR-RIS, UAV-friendly jamming UAV NOMA, UAV_relay_beamforming, encryption, key_generation}. 
AN is that the transmitter in UAV communications transmits a random noise signal together with the valid signal, where the noise is negotiated by the transceivers in advance or is eliminable after transmission~\cite{UAV-friendly jamming STAR-RIS, UAV-friendly jamming UAV NOMA}. 
Precoding is performed by a base station (BS) equipped with multiple antennas or a swarm of UAVs, utilizing the spatial coherence of electromagnetic waves to maximize the reception gain, eliminate interference between receivers, or differentiate legitimate and illegitimate transmissions~\cite{UAV_relay_beamforming}. 
Encryption at the physical layer focuses on key generation and distribution (KDG) by measuring and quantifying the UAV communication channel, and then, transfers the message from plaintext into ciphertext at the symbol level based on the key~\cite{encryption, key_generation}. 
The above techniques have their suitable scenarios and the potential to be employed compatibly for a higher security~\cite{UAV_transmitter_AN_beamforming}. 

On the other hand, physical layer authentication (PLA) can be achieved by using inherent features of legitimate communication links~\cite{authentication_Xie}. 
The device, link, and environment specific features, such as power spectrum density, carrier frequency offset, phase noise, physical unclonable function, received signal strength, channel impulse response, and direction of arrival, can be extracted for authentication purposes~\cite{CSI, CSI2}. 
One challenge of PLA is the lacks of dynamics and quality in these observed physical features. 
Consequently, joint authentication, which considers multiple features with optimized weights, can achieve a higher level of security~\cite{CSI_CFO_PN, PLA_CSI}. 
When the performance of PLA based on available attributes is below requirements, a key-based artificial feature called a tag or fingerprint can be embedded into the message as proof of identity \cite{UAV_conference}. 
A major advantage of this active tag embedding scheme is its more reliable authentication performance when other physical layer features are ineffective~\cite{Paul_part1, Paul_part2}. 
In summary, PLA is an effective technique for ensuring access security in UAV communications, where authentication-referenced features can be flexibly selected to adapt to various scenarios. 

There are several latest and high-related work on security issues in UAV communications. 
In \cite{CSI_CFO_PN}, joint carrier frequency offset and phase noise were utilized to verify the legitimacy of transmitter. 
In \cite{spatial correlation property}, a novel channel sparsity-based authentication scheme was proposed to safeguard communications with multipath channel. 
In \cite{UAV}, tag-based authentication assisted with AN and precoding was proposed to achieve comprehensive security with secrecy and authentication. 
In addition, a serious of cryptography-based approach involving joint encryption, authentication, and key generation through blockchain technology were proposed in \cite{blockchain1,blockchain2,blockchanin3}. 
However, the above-mentioned techniques are often unable to meet the security requirements of UAV communications due to several reasons. 
First of all, in a chaotic UAV communication environment, UAV adversaries become more threatening, employing comprehensive strategies of eavesdropping and jamming, leveraging their numerical and positional advantages, as well as favorable open access conditions. 
The channels of UAV communications are dominated by the line of sight (LoS) paths with sparse scattering~\cite{UAV channel, UAV,CSI_CFO_PN}. 
Such spatial knowledge is easy to be observed and predicated, making UAVs susceptible to imitation and contributing imitators' channels similar to legitimate channels~\cite{risk}. 
Channel similarity could lead to significant performance degradation in AN, precoding, KDG, and channel-based PLA techniques~\cite{encryption2,spatial correlation property}. 
Besides, authentication in downlink communications is more challenging because UAVs do not have sufficient sensing capabilities like the BS to capture recognizable identity features. 
In comparison, tag-based scheme can achieve more reliable authentication due to its independence from the observed physical layer attributes~\cite{tag_Ning_Xie,tag_RIS}. 
Specifically, a tag can be generated by a one-way non-linear mapping of the key and the message, similar to the  message authentication code (MAC) in the upper layer blockchain-based authentication, but MAC is required to be completely correct while the tag can be with tolerable errors~\cite{blockchain1,blockchain2}. 
Thus, the tag embedding scheme is more effective and suitable for resource-constrained UAV communications. 
To achieve the secrecy in UAV communications, encryption can be integrated with the tag-based authentication~\cite{tag_RIS,NOMA,encryption_key,encryption_operation}. 
However, ideal one-time pad encryption requiring high key generation rate is difficult to be implemented in UAV communications~\cite{encryption_key}. 
Also, sufficient complexity for encryption operation is required to defense against differential cryptanalysis, indicating additional significant overhead and computing cost~\cite{encryption_operation}. 
Facing the above computing and physical feature-constrained challenges, the tag for authentication can enhance secrecy at a low cost. 
Due to the differences in tag reception and recovery capability between legitimate receivers and malicious imitators, the tag has the potential to serve as a reference to encode the message from plaintext into ciphertext. 
This idea of integrating secrecy and authentication is novel to solve the security challenges currently faced in UAV communications. 
	
To address the security challenges posed by multiple UAV adversaries at vantage locations with eavesdropping and jamming capabilities, we propose a joint secrecy rate achieving and authentication enhancement scheme for a massive MIMO UAV downlink system. 
Firstly, the tag embedding technique is adopted to guarantee reliable authentication. 
A new tag generation mechanism is proposed by inputting a message's encoding-insensitive feature instead of the message itself (plaintext), along with the key into a hash function. 
This mechanism enables an expected tag regenerable specifically based on the stable feature extracted from the received ciphertext, which can serve as the reference for authentication and decoding. 
Reliable authentication can be performed through a binary hypothesis test by comparing this expected tag with the received one. 
The authenticated ciphertext can then be decoded back into plaintext relying on the expected tag, which remains inaccessible to adversaries and accurate for decryption after authentication. 
Since the tag plays an increasingly important role in this integrated authentication and secrecy security approach, the tag needs to be protected as much as possible. 
In some scenarios with significant channel difference, AN is effective to degrade the wiretap on tag, thereby reducing information leakage caused by decoding attempts based on the wiretapped tag at eavesdroppers.

The main contributions are listed as follows: 
\begin{itemize}
	\item A novel tag-based scheme jointly achieving secrecy rate and enhancing authentication scheme in a massive MIMO UAV downlink system, in the presence of numerous UAV adversaries, is introduced in detail and thoroughly explained. 
	The cores of our proposed approach, including the special-designed dual-reference symmetric tag generation mechanism, tag-based encoding (TBE) and authentication procedures, and signal processing and detection framework, are clearly demonstrated. 
\item 
	Security evaluations are derived and analyzed theoretically to evaluate the security performance of our proposed scheme. 
	Specifically, authentication performance is demonstrated by receiver operating characteristic (ROC), which consists of the false alarm probability and the detection (authentication) probability. 
	Secrecy capability is investigated by ergodic sum secrecy rate, and in addition, a modified expression considering authentication probability and information leakage due to wiretapped tag-based decoding is proposed to assess our method. 
 \item 
 	To achieve effective security and efficiently allocate the power for the tag and AN,  two optimization problems are formulated, each pursuing effectiveness (data rate) and reliability (error ratio), respectively. 
 	One problem focuses on unconstrained ergodic sum secrecy rate maximization, while the second addresses a constrained authentication (transmission) fail probability (AFP) minimization under the constraints of secrecy and authentication. 
 	Additionally,, the difference-of-convex (DC) programming algorithm and the Lagrange method are utilized to solve these problems, respectively. 
\item 
	Simulations based on Monte Carlo method are performed to illustrate the effectiveness of our proposed scheme on security enhancing and its superiority comparing to the prior scheme. 
	No doubt, the simulations align closely with theoretical analyses, also confirming the capability of authentication enhancing and secrecy rate achieving. 
	Furthermore, compared to the prior AN-aided tag embedding scheme, our scheme overcomes the limitation of no secrecy rate when UAV eavesdroppers are positioned in the same direction as legitimate UAVs but closer to the BS. 
	In other scenarios, such as when UAV eavesdroppers are only closer to the BS, the secrecy rate can be improved by 25.8$\%$ and 28.7$\%$, respectively. 
	Additionally, our method achieves a lower AFP compared to the prior non-TBE scheme. 
\end{itemize}

The rest of this paper is organized as follows: Section \ref{system model} presents the system model of communications. 
Section \ref{signal design} firstly introduces the proposed security mechanism and security issues. 
Section \ref{performance analysis} analyzes secrecy and authentication performance theoretically. 
Section \ref{simulation} gives the numerical simulation results and the corresponding analysis. 
Finally, the conclusions are drawn in Section \ref{conclusion}. 
In addition, several important proofs are presented in the Appendix. 

Notation: Scalars, vectors, and matrices are denoted by lowercase italics, lowercase bold, and uppercase bold letters (e.g. $x$, $\mathbf{x}$, and $\mathbf{X}$), respectively. 
Vectors are assumed to be column vectors. 
Subscripts in normal font distinguish different objects (e.g. t, m, s, and n), and subscripts in italic font are variables (e.g. $u$ and $e$).
Decorated letters denote binary strings or constellation sets (e.g. $\mathcal{M}$ and $\mathcal{X}$). 
$\mathbf{I}$ represents the identity matrix, and $\mathbf{0}$ is a zero vector. 
$\left(\cdot\right)^T$ and $(\cdot)^H$ denote transpose and conjugate transpose, respectively. 
$|\cdot|$ denotes the absolute value, while $\|\cdot\|$ denotes the norm. 
$\oplus$ is the XOR operator, and $\odot$ is the Hadamard product. 
$<\cdot,\cdot>$ is the inner product. 
$Q(\cdot)$ denotes the Q-function, and $C^{a}_{b}$ denotes the combination operator. 
$\mathcal{CN}(\cdot, \cdot)$ and $\mathcal{U}(\cdot, \cdot)$ denote complex normal and uniform distributions, respectively, where the first input is the mean and the second is the variance or covariance matrix. 
The hash function is denoted by $\text{hash}(\cdot)$. 
The sum function is denoted by $\text{sum}(\cdot)$. 
The supremum operator is denoted by sup$\left\{\cdot\right\}$. 
$\mathbb{E}(\cdot)$ denotes the expectation. 
$\exp(\cdot)$ denote the natural exponential function. 
$\left(\cdot\right)^+$ returns the maximum of the input and zero. 
The imaginary unit is denoted by $j = \sqrt{-1}$. $\mathcal{R}(\cdot)$ and $\mathcal{I}(\cdot)$ return the real and imaginary parts of the input, respectively. 
$\log_2(\cdot)$ and $\lg(\cdot)$ denote the base-2 and base-10 logarithms, respectively.

\begin{table}[!b]
	\caption{Key notations and evaluations. }
	{\begin{tabular*}{20pc}{@{\extracolsep{\fill}}|c|c|@{}}\hline
			Notation  &Meaning \\
			\hline
			$\phi_{\rm s}/\phi_{\rm n}$ ($\phi_{\rm s}^2+\phi_{\rm n}^2=1$)& Artificial noise power allocation\\
			\hline
			$\rho_{\rm m}/\rho_{\rm t}$ ($\rho_{\rm m}^2+\rho_{\rm t}^2=1$)& Tag embedding power allocation\\
			\hline
			$\text{SINR}_{u,\rm{m/t}}/\text{SINR}_{e,\rm{m/t}}^{'}$&Signal-to-interference-plus-noise ratio      \\ 
			\hline
			$P_{u,\rm{m/t}}/P_{e,\rm{m/t}}^{'}$& Symbol error ratio (SER)\\
			\hline
			$P_{u,\rm{d}}/P_{\rm{d}}$&Detection (authentication) probability \\ 
			\hline 
			$P_{u,\rm{f}}/P_{\rm{f}}$&False alarm probability\\
			\hline
			$R_{\rm{U}}/R_{\rm{E}}$&Users'/EVEs' data rate \\
			\hline
			$R_{\rm{sec}}$& Ergodic sum secrecy rate\\
			\hline
			$\text{AFP}_u$/AFP&Authentication fail probability 
			\\
			\hline
			$P_{\rm{w}}$& Wiretap SER
			\\
			\hline
			$\rho=\rho_{\rm m}^2$, $\phi=\phi_{\rm s}^2$& Optimization variables
			\\
			\hline
	\end{tabular*}}{}\label{Notations}
\end{table}
For clarity, key notations and evaluations are summarized in Tab. \ref{Notations}. 
Specifically, subscripts $u$ and $e$ are the indices of users and eavesdroppers, respectively. 
Subscripts $\rm{m}$ and $\rm{t}$ refer to the message and the tag, respectively.  
The evaluations of users and eavesdroppers are distinguished by the absence or presence of a superscript
$'$, with $'$ indicating terms related to eavesdroppers.
Notations with a hat ($\hat{\cdot}$) represent estimated expressions, whereas those with a tilde ($\tilde{\cdot}$) denote derived intermediate or regenerated expressions, such as $\mathbf y/\hat{\mathbf{y}}/\tilde{\mathbf{y}}$ and $\mathbf t/\hat{\mathbf{t}}/\tilde{\mathbf{t}}$. 
Evaluations without an index signify average values; for example, $P_{\rm{d}}$ denotes the average of $P_{u,\rm{d}}$.

\section{System Model}
\label{system model}
\begin{figure}[!t]\centering
	\includegraphics[width=3.4in]{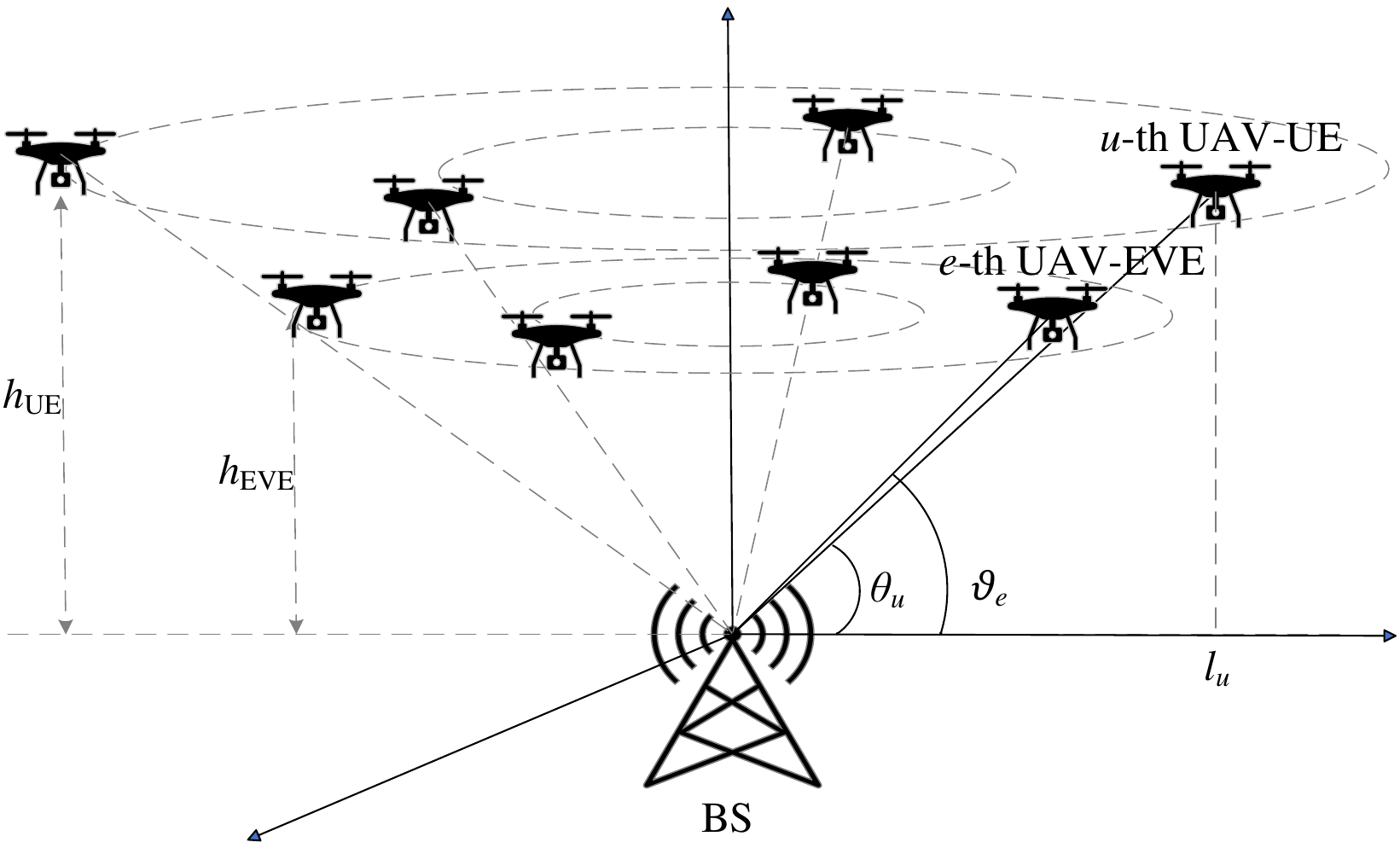}
	\caption{Downlink communications from BS toward multiple UAV users with the presence of UAV adversaries. }\label{model}
\end{figure}
As shown in Fig. \ref{model}, we consider a chaotic UAV communication scenario where messages are transmitted block by block from a BS equipped with $M$ antennas toward $K$ UAV users (UAV-UEs) each equipped with a single antenna in the presence of numerous UAV adversaries, where each block contains $T$ time slots. 
The UAV-UEs fly at the same altitude $h_{\rm{UE}}$ with different elevation angles $(\theta_1, \cdots, \theta_K)$. 
There are $K$ UAV eavesdroppers (UAV-EVEs) deployed at vantage locations between the BS and UAV-UEs. 
\footnote{We set $K$ eavesdroppers to simplify analysis. 
The increasing of eavesdroppers indicates a higher security risk and can be analyzed by modifying the reception at UAV-EVEs.  }
Their flying altitudes are the same $h_{\rm{EVE}}$ and their elevation angles ($\vartheta_1, \cdots, \vartheta_K$) are close to those of UAV-UEs. 
They are connected via a central processing unit (CPU) and attempt to wiretap messages sent by the BS and jam users with false messages via replay or imitation. 
In order to protect UAV communications, basic defenses including ZF precoder and null-space AN are established to upgrade legitimate transmissions and degrade illegitimate transmissions.

The signal received at $u$-th UAV-UE ($u=1,\cdots,K$ and it is substituted by index variable $k$ in summation) is formulated as 
\begin{equation}
	\mathbf{y}_u^H=\sqrt{{P_{\rm T}}\beta_u}\mathbf{h}_u^H\left(\phi_{\rm s}\sum\limits_{k = 1}^{{K}}\mathbf{w}_k\mathbf{x}_k^H
	+ \phi_{\rm n}\sum\limits_{i = 1}^{{N_{\rm{AN}}}} {{\mathbf v_i}\mathbf z_i^H}\right)+\mathbf n_u^H,
\end{equation}
where $\mathbf y_u\in\mathbb{C}^{T*1}$ is the received symbol vector, $P_{\rm T}$ is the transmit power, $\beta_u$ is the large-scale fading coefficient, $\mathbf h_u\in\mathbb{C}^{M*1}$ is the small-scale fading vector, and $\mathbf n_u\in\mathbb{C}^{T*1}$ is the additive white Gaussian noise (AWGN) vector whose entities follow the zero-mean Gaussian distribution with variance $\sigma_{\rm n}^2$. 
Besides, the BS transmits the signal and AN parts, associated by power allocations $\phi_{\rm s}$ and $\phi_{\rm n}$ ($\phi_{\rm s}^2+\phi_{\rm n}^2=1$). 
The signal part contains the precoder $\mathbf{w}_k\in\mathbb{C}^{M*1}$ and the signal vector $\mathbf{x}_k\in\mathbb{C}^{T*1}$. 
The AN part consists of the basis vector $\mathbf v_i\in\mathbb{C}^{M*1}$ and the zero vector $\mathbf z_i\in\mathbb{C}^{T*1}$ ($i$ is the index variable in summation indicating the dimension of AN), where the dimensions of AN are $N_{\rm{AN}}$ and all entities of $\mathbf z_i$ follow the zero-mean unit variance complex Gaussian distribution. 
In addition, the ZF preorder for $u$-th UAV-UE $\mathbf w_u$ and the AN design will be given in Section \ref{ZF_AN}, and the signal processing for $\mathbf x_u$ is specifically described in Section \ref{Tag embedding and signal encoding}.

\subsection{Channel model and UAV's location}\label{Channel}
The channels between the BS and UAVs are modeled as Rician fading channels, which can be expressed as 
\begin{equation}
	\mathbf h=\sqrt{\frac{\kappa}{\kappa+1}}\mathbf h_{\rm{LoS}}+\sqrt{\frac{1}{\kappa+1}}\mathbf h_{\rm{NLoS}},\label{channel}
\end{equation}
where $\mathbf{h}_{\rm{LoS}}$ is the LoS component, while $\mathbf{h}_{\rm{NLoS}}$ is the scattered random non-line of sight (NLoS) component. 
These two components are characterized by the Rician $\kappa$-factor. 
Assuming that the BS is equipped with vertically oriented uniform linear array (ULA) antennas, the LoS component can be represented as  
\begin{equation}
	\mathbf h_{\rm{LoS}}=\left[1~ e^{-j\frac{2\pi d_{\rm s}}{\lambda}\sin\theta}\cdots e^{-j\frac{2\pi d_{\rm s}}{\lambda}\left(M-1\right)\sin\theta}\right]^T,\label{LoS}
\end{equation}
where $\lambda$ is the wavelength, $d_{\rm s}$ is antennas spacing (typically set to $\lambda/2$), and $\theta$ is the elevation angle. 
The NLoS paths are full of randomness and its elements follow the complex normal distribution, i.e., $\mathbf{h}_{\rm{NLoS}} \sim \mathcal{CN}(\mathbf{0}_M, \mathbf{I}_M)$.

The large-scale coefficient can be calculated from the path loss in the Urban Micro (UMi, 3GPP) scenario as 
\begin{equation}
\beta=1/\left(32.4+21\lg\left(d\right)+20\lg\left(f_{\rm c}\right)\right),\label{beta}
\end{equation}
where $d$ is the distance between BS and UAV while $f_{\rm c}$ is the normalized carrier frequency ($0.5~\text{GHz}<f_{\rm c}<100~\text{GHz}$). 

Both $\mathbf h_{\rm{LoS}}$ and $\beta$ are determined by UAV-UEs' locations. 
We assume that UAV-UEs are deployed randomly around the BS. 
The horizontal distance ($l$) between the BS and UAV-UE follows a uniform distribution: $l\sim\mathcal{U}(l_{\text{min}}, l_{\text{max}})$.
Parameters $\theta$ in (\ref{LoS}) and $d$ in (\ref{beta}) can be computed by $\theta=\arctan\left({h_{\text{UE}}}/{d}\right)$ and $d=\sqrt{l^2+h_{\text{UE}}^2}$, respectively. 

\subsection{ZF precoder and null-space AN}
\label{ZF_AN}
ZF precoder aims to eliminate the interference between UAV-UEs, while null-space AN can degrade the wiretap channels. 
Both of them are calculated based on the channel matrix 
$\mathbf H=\left[\mathbf h_1,\cdots,\mathbf h_K\right]$.
Particularly, we assume $\mathbf H$ has been estimated perfectly because the channels are mainly determined by the LoS paths. 
The unnormalized precoder matrix $\tilde{\mathbf W} = \left[\tilde{\mathbf w}_1, \ldots, \tilde{\mathbf w}_K\right]$ can be expressed as
\begin{equation}
  \tilde{\mathbf W}=\mathbf H\left(\mathbf H^H\mathbf H\right)^{-1}.
\end{equation}
After users' transmission balance and power normalization, the ZF precoder for $u$-th UAV-UE is formed as 
\begin{equation}
	\mathbf w_u=\sqrt{\frac{\beta_u^{-1}}{\sum\nolimits_{k = 1}^{{K}}\beta_k^{-1}}}\frac{\tilde{\mathbf w}_u}{\|\tilde{\mathbf w}_u\|}.\label{precoder}
\end{equation}
So far, we have $\sum\nolimits_{k = 1}^K \|\mathbf{w}_k\|^2=1$, $\beta_u\|\mathbf{w}_u\|^2=1/{\sum\nolimits_{k = 1}^{{K}}\beta_k^{-1}}$, $\|\mathbf h_u^H\mathbf{w}_k\|^2|_{k\ne u}=0$, and $\|\mathbf h_u^H\mathbf w_u\|^2=\beta_u^{-1} M/{\sum\nolimits_{k = 1}^{{K}}\beta_k^{-1}}$. 

The null-space AN basis vector $\mathbf v$ can be obtained by solving $\mathbf{H}^H\mathbf v=\mathbf{0}$. 
There are $N_{\rm{AN}}=M-K$ solutions, denoted by $\tilde{\mathbf v}_i$ ($i=1,\cdots,N_{\rm{AN}}$). 
After balancing the norm of the solutions and normalizing the total power, the AN basis vector can be given as 
\begin{equation}
	\mathbf v_i=\frac{\tilde{\mathbf v}_i}{N_{\rm{AN}}\|\tilde{\mathbf v}_i\|}.\label{AN}
\end{equation}

\section{The proposed Tag-based authentication and encoding}\label{signal design}
Signal processing procedures contain modulation, detection, and security mechanisms.
The security mechanisms indicate the tag embedding and message encoding at the BS, and the corresponding authentication and decoding procedures. 
All transmission procedures are presented mathematically in this section.

\subsection{Tag embedding and signal encoding}
\label{Tag embedding and signal encoding}
At the BS, for $u$-th UAV-UE, the message string and the modulated message symbol vector are denoted by $\mathcal{M}_u$ and $\mathbf s_u\in\mathbb{C}^{T*1}$, respectively. 
Here, assuming that $2^n$-PSK ($n\ge 2$) modulation is adopted with the message mapping constellation $\mathcal{X}_{\rm m}:\left\{\exp{\left(2\pi\left(\frac{c}{2^n}+\frac{1}{2^{n+1}}\right)\right)}|c=0,\cdots,2^n-1\right\}$, the length of $\mathcal{M}_u$ is set to $nT$. 
The secret key $\mathcal{K}$ of sufficient length has been shared through secure transmission.

For security, a tag $\mathcal{T}_u$ ($T$ bits) is generated to be embedded into $\mathbf{s}_u$. 
In prior, the tag was only embedded to certify the legitimacy for authentication, so it is generated by inputting the key and the message into the hash function. 
In our design, the tag is also utilized to encode the message for secrecy at a low cost. 
Thus, the tag generation input is modified as the key together with a message feature $\mathcal{F}_u$. 
The above dual-reference symmetric tag generation modification can be demonstrated as 
\begin{equation}
    \mathcal T_u=\text{hash}\left(\mathcal{K},\mathcal M_u\right) \to \mathcal T_u=\text{hash}\left(\mathcal{K},\mathcal F_u\right). 
\end{equation}
In addition, the feature and encoding manner require a specific design. 
On one hand, the feature should be sensitive to the message, that is, it changes when the message changes. 
On the other hand, the feature should be stable for encoding. 
The sensitivity results from authentication accuracy requirements, while the encoding-unchangeable property makes decoding possible. 

A paradigm is given as follows. 
In each slot, if the symbol is located in the first or third quadrant, the corresponding bit of the feature is set to $1$, otherwise, it is set to $0$. 
Mathematically, this feature extraction can be conveyed as 
\begin{equation}
	\mathcal F_{u}=\mathcal R\left(\mathbf s_u\right)\oplus \mathcal I\left(\mathbf s_u\right).\label{feature}
\end{equation}
Correspondingly, a matching encoding manner is phase reversal. 
In $\tau$-th ($\tau=1,\cdots, T$) slot, the symbol $s_{u,\tau}$ is transferred into $-s_{u,\tau}$ if $\mathcal{T}_{u,\tau}$ is $0$, otherwise, the symbol remains unchanged ($s_{u,\tau}$ and $\mathcal T_{u,\tau}$ are $\tau$-th elements of $\mathbf s_u$ and $\mathcal T_u$, respectively). 
It is worth noting that the feature extraction process, which carries out the XOR of the real and imaginary parts of the message, is lightweight. Furthermore, compared to the inherent hash mapping calculations, the additional overhead introduced by feature extraction is minimal and can be safely disregarded. Therefore, the overhead introduced by our newly proposed dual-reference symmetric tag generation is negligible. 

As a result, the transmitted signal $\mathbf x_u$ is processed as 
\begin{equation}
	\mathbf{x}_u=\rho_{\rm m}\mathbf t_u\odot\mathbf s_u+\rho_{\rm t}\mathbf t_u,\label{signal}
\end{equation}
where 
$\mathbf t_u$ is the tag symbol vector modulated by $\mathcal{X}_{\rm t}:\left\{-1, 1\right\}$ from $\mathcal T_u$, and $\rho_{\rm m}$ and $\rho_{\rm t}$ are the power allocated to the message and the tag ($\rho_{\rm m}^2+\rho_{\rm t}^2=1$ and $\rho_{\rm m}\gg \rho_{\rm t}$), respectively. 
Besides, the encoded symbol vector as well as the ciphertext, is denoted by $\mathbf c_u=\mathbf t_u\odot \mathbf s_u$. 

\subsection{Signal detection and authentication}
\label{signal authentication and decoding}
At the receiver ($u$-th UAV-UE), the message is detected at first by treating the tag as noise. 
Let $\tilde{\mathbf y}_u=\mathbf{y}_u/\phi_{\rm s}\sqrt{P_{\rm T}\tilde{\beta}M}$, where $\tilde{\beta}=1/\sum\nolimits_{k = 1}^{K}\beta_k^{-1}$. 
Based on the minimum Euclidean distance criterion, $\mathbf{c}_u$ can be detected slot by slot, and its entity is obtained by 
\begin{equation}
	\hat{c}_{u,\tau}= \mathop {\arg \min }\limits_{s \in\mathcal{X}_{\rm m}} | {\tilde{y}_{u,\tau}-\rho_{\rm m}s} |,\label{message detection}
\end{equation}
where $\hat{c}_{u,\tau}$ and $\tilde{y}_{u,\tau}$ are the $\tau$-th ($\tau=1,\cdots,T$) elements of the ciphertext estimator $\hat{\mathbf c}_{u}$ and $\tilde{\mathbf{y}}_u$, respectively. 

The message will be decoded if it is authenticated. 
Otherwise, it will be discarded, whatever it is transmitted with errors from the BS or is a false message from an adversary.
The authentication relies on the tag embedding mechanism. 
The embedded tag can be detected from the residual signal $\mathbf r_u=\tilde{\mathbf{y}}_u-\rho_{\rm m}\hat{\mathbf c}_{u}$, and the detection is shown as 
\begin{equation}
	\hat{t}_{u,\tau}=\mathop {\arg \min }\limits_{t \in\mathcal{X}_{\rm t}} | {{r}_{u,\tau}-\rho_{\rm t}t} |,\label{tag detection}
\end{equation}
where $\hat{t}_{u,\tau}$ and ${r}_{u,\tau}$ are the $\tau$-th elements of the detected tag $\hat{\mathbf t}_{u}$ and ${\mathbf{r}}_u$, respectively. 
Moreover, an expected tag can be regenerated according to the received ciphertext $\hat{\mathbf{c}}_u$, shown as 
\begin{equation}
	\tilde{\mathbf {\mathcal{T}}}_u=\text{hash}\left(\mathcal{K},\hat{\mathcal{F}}_u\right),
\end{equation}
where $\hat{\mathcal{F}}_u=\mathcal{R}\left(\hat{\mathbf c}_{u}\right)\oplus\mathcal{I}\left(\hat{\mathbf c}_{u}\right)$ is similar to (\ref{feature}). 
Then, $\tilde{\mathcal{T}}_u$ can be modulated as $\tilde{\mathbf{t}}_u$ using the constellation $\mathcal X_{\rm t}$. 

Specifically, authentication is created as a binary hypothesis test. 
The two hypotheses are listed as 
\begin{itemize}
	\item[] $\mathcal H_0$: the message is illegitimate, not authentic, 
	\item[] $\mathcal H_1$: the message is legitimate, authentic. 
\end{itemize}
The test is constructed as 
\begin{equation}
	\mathcal{L}\left(\hat{\mathbf t}_u,\tilde{\mathbf t}_u\right)=\text{sum}\left(\hat{\mathbf t}_u\oplus\tilde{\mathbf t}_u\right)\mathop {\buildrel>\over
		{\smash{\scriptstyle<}\vphantom{_x}}}\limits_{\mathcal H_1}^{\mathcal H_0}\eta,\label{statistic}
\end{equation}
where the statistic $\mathcal{L}$ indicates the error bits between the detected tag $\hat{\mathbf t}_u$ and the regenerated tag $\tilde{\mathbf t}_u$, and $\eta$ is the threshold that determines the security level. 
The hypothesis $\mathcal{H}_0$ holds if $\mathcal{L}$ is greater than $\eta$, otherwise, $\mathcal{H}_1$ holds.

Once the message is authenticated, the regenerated tag $\Tilde{\mathbf t}_u$ is reliable to decode the ciphertext $\hat{\mathbf c}_{u}$ as plaintext $\hat{\mathbf s}_u$. 
The decoding process can be expressed as $\hat{\mathbf s}_u=\tilde{\mathbf t}_u\odot \hat{\mathbf c}_u$. 
At last, the message is demodulated as $\hat{\mathcal{M}}_u$. 

\subsection{Security issues}
There are $K$ UAV-EVEs strategically deployed in favorable locations to individually threaten each UAV-UE through eavesdropping and jamming. 
Favorable locations indicate a lower height $h_{{\rm{{EVE}}}}\le h_{\rm{UE}}$ and similar elevation angles. 
For simplicity, we assume that the difference between elevation angles is the same, i.e., $\vartheta_e-\theta_u|_{u=e}=\Delta\theta$. 
This difference is caused by detection errors and flying disturbances. 
According to the channel model provided in Section \ref{Channel}, the wiretap channels can be modeled as a matrix $\mathbf G=\left[\mathbf g_1,\cdots,\mathbf g_K\right]$, where $\mathbf g_e\in\mathbb{C}^{M*1}$ ($e=1,\cdots,K$) is the wiretap channel between the $e$-th UAV-EVE and the BS. 
In detail, $\mathbf{g}_e$ contains the LoS component determined by $h_{\rm{EVE}}$ and $\vartheta_e$ and the NLoS component full of randomness. 
Besides, the large-scale coefficient $\alpha_e$ can be calculated based on the $e$-th UAV-EVE's location according to (\ref{beta}). 
Thus, the wiretap signal matrix $\mathbf Y_{\rm E}$ is written as 
\begin{equation}
\mathbf{Y}_{\rm E}^H=\sqrt{P_{\rm T}}\mathbf{D}_{\rm E}^{1/2}\mathbf{G}^H\left(\phi_{\rm s}\mathbf{W}\mathbf{X}^H+\phi_{\rm n}\mathbf{V}\mathbf{Z}^H\right)+\mathbf N_{\rm E}^H,
\end{equation}
where $\mathbf D_{\rm E}=\text{diag}\left\{\alpha_1,\cdots,\alpha_K\right\}$, $\mathbf{Z}=\left[\mathbf z_1, \cdots, \mathbf z_{N_{\rm{AN}}}\right]$ is the zero matrix component whose elements follow the zero-mean unit variance complex Gaussian distribution, and $\mathbf N_{\rm E}\in\mathbb{C}^{K*T}$ is the noise matrix whose elements follow the zero-mean complex Gaussian distribution with variance $\sigma_{\rm n}^2$. 
In addition, $\mathbf W=\left[\mathbf w_1, \cdots, \mathbf w_K\right]$, $\mathbf V=\left[\mathbf v_1, \cdots, \mathbf v_{N_{\rm{AN}}}\right]$, and $\mathbf X=\left[\mathbf{x}_1, \cdots, \mathbf{x}_K\right]$ can be obtained according to (\ref{precoder}), (\ref{AN}), and (\ref{signal}), respectively. 

We assume that adversaries are powerful enough to measure the channels precisely. 
Correspondingly, UAV-EVEs are observed by the BS, i.e., LoS components of illegitimate channels are known at the BS. 
The CPU has the knowledge about the security mechanism, legitimate and illegitimate channels, including $\mathbf{D}_{\rm E}$, $\mathbf{G}$, $\mathbf{H}$, $\mathbf{W}$, $\phi_{\rm s}$, $\phi_{\rm n}$, $\rho_{\rm m}$, and $\rho_{\rm t}$. 
The normalized received signal $\tilde{\mathbf{Y}}_{\rm E}$ can be expressed as
\begin{equation}
\tilde{\mathbf{Y}}_{\rm E}=\sqrt{1/\phi_{\rm s}P_{\rm T}}\mathbf{Y}_{\rm E}\mathbf{D}_{\rm E}^{-1/2}\mathbf{G}^H\mathbf{W}\left(\mathbf{W}^H\mathbf{G}\mathbf{G}^H\mathbf{W}\right)^{-1}. 
\end{equation}
When it comes to the message and the tag symbol recovery, the detection methods given in (\ref{message detection}) and (\ref{tag detection}) are considerable, and in addition, the wiretapped message and tag matrices are denoted by $\hat{\mathbf C}_{\rm E}$ and $\hat{\mathbf T}_{\rm E}$.

Adversaries do not have access to the key, thus they are unable to regenerate the expected tags for decoding. 
To capture as much information as possible, adversaries may attempt to decode the ciphertext using the wiretapped tag matrix, i.e., $\hat{\mathbf{S}}_{\rm E} = \hat{\mathbf{C}}_{\rm E} \odot \hat{\mathbf{T}}_{\rm E}$. 
The wiretap performance is primarily determined by the accuracy of $\hat{\mathbf{C}}_{\rm E}$, $\hat{\mathbf{T}}_{\rm E}$, and $\hat{\mathbf{S}}_{\rm E}$. Generally, the replay attack can be easily detected by comparing the current tag to previous tags~\cite{authentication_Xie}. 
Therefore, the jamming method is considered to transmit random messages and tags. 

\section{Performance Analysis}\label{performance analysis}
In this section, the communication performance is analyzed by mathematically deriving the closed-form expressions of SINRs and SERs.
Furthermore, the authentication performance at UAV-UEs is demonstrated by authentication probability and false alarm probability. 
Besides, the security issues are concluded as an unconstrained ergodic sum secrecy rate maximization problem and an authentication (transmission) fail probability minimization problem under the constraints of the authentication probability and the wiretap SER. 
Additionally, the difference-of-convex (DC) algorithm and the Lagrange method are utilized to solve these two optimization problems for power allocations. 

\subsection{Legitimate transmission}
At UAV-UEs, the SINRs of messages and tags are derived first. 
The interference between users is eliminated by ZF precoding while null-space AN does not introduce interference to signals. 
The SINR of message at $u$-th UAV-UE is given as 
\begin{equation}
	\begin{aligned}
		&\text{SINR}_{u,\rm{m}}= \\& \mathbb{E}\left\{\frac{{P_{\rm T}\beta_u\phi_{\rm s}^2\rho_{\rm m}^2{\bf{w}}_u^H{{\bf{h}}_u}{\bf{h}}_u^H{{\bf{w}}_u}}}{{P_{\rm T}\beta_u\phi_{\rm s}^2\rho _{\rm t}^2{\bf{w}}^H{{\bf{h}}_u}{\bf{h}}_u^H{{\bf{w}}} + \sigma_{\rm n}^2}}\right\}= \frac{{\phi_{\rm s}^2\rho _{\rm m}^2M}}{{\phi_{\rm s}^2\rho_{\rm t}^2M + \sigma_{\rm n}^2/{P_{\rm T}\tilde{\beta}}}},
	\end{aligned}\label{{SINR}_{u,c,AN}}
\end{equation}
where the interference comes from the tag. 
The SINR of tag at $u$-th UAV-UE can be presented as 
\begin{equation}
	\text{SINR}_{u,\rm{t}} = \mathbb{E}\left\{ {\frac{{P_{\rm T}\beta_u\phi_{\rm s}^2\rho_{\rm t}^2{\bf{w}}_u^H{{\bf{h}}_u}{\bf{h}}_u^H{{\bf{w}}_u}}}{{\sigma_{\rm n}^2/2}}} \right\} = \frac{{2\phi_{\rm s}^2\rho_{\rm t}^2M}}{{\sigma_{\rm n}^2/P_{\rm T}\tilde{\beta}}}.\label{{SINR}_{u,t,AN}}
\end{equation}

For analyzing the transmission performance at UAV-UEs from a probabilistic perspective, SERs of message and tag are investigated, with the modulation order set to $n = 2$ (QPSK). 
The SER of the ciphertext message is expressed as 
\begin{equation}
	\begin{aligned}
		&{P_{u,\rm{m}}} =\\&1- \frac{1}{2}{\left( {Q\left( {\frac{{\frac{\rho_{\rm m}}{\sqrt 2 } + {\rho_{\rm t}}}}{{ - {\sigma _{u}}}}} \right) + Q\left( {\frac{{\frac{{{\rho_{\rm m}}}}{{\sqrt 2 }} - {\rho_{\rm t}}}}{{ - {\sigma _{u}}}}} \right)} \right)Q\left( {\frac{{\frac{{{\rho_{\rm m}}}}{{\sqrt 2 }}}}{{ - {\sigma _{u}}}}} \right)},
	\end{aligned}
	\label{P_s_u}
\end{equation}
where $Q(x)=\int_x^{\infty }\frac{1}{\sqrt{2\pi}}{e^{-\frac{\xi^2}{2}}}{d\xi}$ is the Q-function and we have 
\begin{equation}
	\sigma_{u}=\sigma_{\rm n}/\phi_{\rm s}\sqrt{2MP_{\rm T}\tilde{\beta}}.
\end{equation}
Specifically, $\sigma_u$ is the standard deviation of the equivalent AWGN. 
(\ref{P_s_u}) presents the SER of the QPSK symbol considering the interference caused by the tag embedding, where the probabilities of tag symbol being positive or negative are both $1/2$. 
Obviously, it changes to the standard form of QPSK's SER when $\rho_{\rm t}=0$. 
The SER of tag is related to the detection of message symbol and error cases. 
Assuming that $s+t\to s-t$ indicates the case where the transmitted tag symbol is incorrect while the message symbol is correct. 
The SER of tag is calculated by 
\begin{equation}
	\begin{aligned}
		{P_{u,\rm{t}}} &=\frac{1}{2}\left(\right. \text{Pr}\left\{s+t\to s-t\right\}+\text{Pr}\left\{s-t\to s+t\right\}\\
  &+\text{Pr}\left\{s+t\to -s-t\right\}+\text{Pr}\left\{s-t\to -s+t\right\}\left.\right)\\
		&=\frac{1}{2}\left( {Q\left( {\frac{{{\rho_{\rm t}}}}{\sigma _{u} }} \right) - Q\left( {\frac{{\frac{\rho_{\rm m}}{\sqrt 2}  + {\rho_{\rm t}}}}{\sigma _{u} }} \right) + Q\left( {\frac{{\sqrt 2 {\rho_{\rm m}} + {\rho_{\rm t}}}}{\sigma _{u} }} \right)}\right.\\
				&\left.{ +Q\left( {\frac{{\frac{\rho_{\rm m}}{\sqrt 2}  - {\rho_{\rm t}}}}{\sigma _{u} }} \right) - Q\left( {\frac{{\sqrt 2 {\rho_{\rm m}} - {\rho_{\rm t}}}}{\sigma _{u} }} \right)+Q\left( {\frac{{{\rho_{\rm t}}}}{\sigma _{u} }} \right) } \right).
	\end{aligned}
	\label{P_t_u}
\end{equation}
Under the assumption that $\rho_{\rm m}>>\rho_{\rm t}$, (\ref{P_t_u}) can be simplified as $P_{u,\rm{t}}=Q\left(\rho_{\rm t}/\sigma_u\right)$.

At UAV-UEs, only authenticated messages will be accepted and then decoded. 
The authentication performance is evaluated by authentication (detection) probability, the probability that the transmitted message is correct and the tag's error is smaller than $\eta$. 
Besides, the receptions of message and tag at UAV-UEs have been balanced by the weighted precoder in (\ref{precoder}). 
The authentication probability at $u$-th UAV-UE is calculated as 
\begin{equation}
	{P_{u,\rm d}} = P(\mathcal H_1|\mathcal H_1)= \sum\limits_{\zeta = 0}^\eta  {C_T^{\zeta} \cdot P_{u,\rm{t}}^{\zeta}\cdot {{\left( {1 - {P_{u,\rm{t}}}} \right)}^{T-\zeta}} }, \label{pd}
\end{equation}
where $\zeta$ is the index variable indicating the error bits of the received tag.

Generally, the false alarm probability is the direct representation of the security level. 
It is the likelihood of mistaken authentication. 
Two scenarios can cause false alarms. 
In the first scenario, the transmitted message has errors and leads to changes in its feature, which will cause the regenerated tag to change randomly. 
This significant change indicates that each bit of the regenerated tag has a probability of $0.5$ of matching the embedded one. 
In the second scenario, one or several symbols of message are mistakenly reversed due to transmission errors, which will not cause a change in the estimated feature and the regenerated tag.
This kind of false message may result in a false alarm if the error of the transmitted tag is less than $\eta$. 
Above these two cases, the false alarm probability at $u$-th UAV-UE is expressed as 
\begin{equation}
	\begin{aligned}
		&{P_{u,\rm f}} = P({{\cal H}_1}|{{\cal H}_0})\\
		&=\left(1-P_{u,{\rm b}}\right) \sum\limits_{\zeta = 0}^\eta  {C_T^\zeta  \frac{1}{{{2^T}}}}  + {P_{u,{\rm b}}}  \sum\limits_{\zeta = 0}^\eta  {C_T^\zeta  P_{u,\rm{t}}^\zeta  {{\left( {1 - {P_{u,\rm{t}}}} \right)}^{T - \zeta}}},
	\end{aligned}
	\label{pf}
\end{equation}
where $1-P_{u,\rm b}$ and $P_{u,\rm b}$ are the likelihoods of the first and second scenarios, respectively. 
Based on the SER of message, $P_{u,\rm b}$ can be given as 
\begin{equation}
	{P_{u,\rm b}} = \sum\limits_{\zeta = 1}^T {C_T^\zeta  {{\left( {{{1-P_{u,\rm{m}}}}} \right)}^{T-\zeta}}{{\left( {1 - \sqrt {{1-P_{u,\rm{m}}}} } \right)}^{2\zeta }}}.
\end{equation}
It is worth mentioning that the second term in (\ref{pf}) is an additive increment introduced by the tag generation modification. 
This increment contributes to a higher $P_{u,\rm f}$, but it is much smaller than the first inherent term and its impact on $P_{u,\rm f}$ is negligible. 
Due to the benefits of enabling tag-based encoding and decoding for secrecy, this limited degradation in authentication is acceptable for comprehensive security. 

\subsection{Wiretap analysis}\label{wiretap analysis}
For simplicity, we assume that the interference between UAV-EVEs can be eliminated using successive interference cancellation (SIC). 
In other words, each UAV-EVE is assigned to wiretap one specific UAV-UE and is not interfered with by other UAV-UEs ($u=e$ when they appear at the same time). 
The SINR of message at $e$-th UAV-EVE is written as 
\begin{equation}
	\begin{aligned}
		&\text{SINR}_{e,\rm{m}}^{'}= \\ &
		\mathbb{E}\left\{ {\frac{{\phi_{\rm s}^2\rho_{\rm m}^2{\bf{w}}_u^H{{\bf{g}}_e}{\bf{g}}_e^H{{\bf{w}}_u}}}{{\phi_{\rm s}^2\rho_{\rm t}^2{\bf{w}}_u^H{{\bf{g}}_e}{\bf{g}}_e^H{{\bf{w}}_u} + \phi_{\rm n}^2\sum\limits_{i = 1}^{{N_{NA}}} {\mathbf v_i^H{{\bf{g}}_e}{\bf{g}}_e^H{\mathbf v_i}}  + {\sigma_{\rm n}^2}/{P_{\rm T}\alpha_e}}}} \right\},
	\end{aligned}\label{{SINR}_{e,c,AN}}
\end{equation}
where the superscript ``$\prime$'' is committed to distinguishing UAV-EVEs' evaluations from those of UAV-UEs.
Firstly, we have  $\mathbb{E}\left\{{{{\bf{w}}_u^H{{\bf{g}}_e}{\bf{g}}_e^H{{\bf{w}}_u}}}\right\}=\mathbb{E}\left\{{\bf{h}}_u^H{{\bf{g}}_e}{\bf{g}}_e^H{{\bf{h}}_u}\right\}\beta_u^{-1}\tilde{\beta}/M=\beta_u^{-1}\tilde{\beta}f\left(\kappa\right)$. 
Dividing the channel vectors into LoS and NLoS components, we have 
\begin{equation}
\begin{aligned}
    &f\left(\kappa\right)={\kappa^2}/{M\left(\kappa+1\right)^2}\cdot\mathbf{h}_{u,\rm{LoS}}^H{{\mathbf{g}}_{e,\rm{LoS}}}{\mathbf{g}}_{e,\rm{LoS}}^H{{\mathbf{h}}_{u,\rm{LoS}}}
    \\
    &+{\kappa}/{M\left(1+\kappa\right)^2}\cdot\mathbb{E}\left\{{\bf{h}}_{u,\rm{LoS}}^H{{\bf{g}}_{e,\rm{NLoS}}}{\bf{g}}_{e,\rm{NLoS}}^H{{\bf{h}}_{u,\rm{LoS}}}\right\}
    \\
    &+{\kappa}/{M\left(1+\kappa\right)^2}\cdot\mathbb{E}\left\{{\bf{h}}_{u,\rm{NLoS}}^H{{\bf{g}}_{e,\rm{LoS}}}{\bf{g}}_{e,\rm{LoS}}^H{{\bf{h}}_{u,\rm{NLoS}}}\right\}
    \\
    &+{1}/{M\left(1+\kappa\right)^2}\cdot\mathbb{E}\left\{{{\bf{h}}_{u,\rm{NLoS}}^H{{\bf{g}}_{e,\rm{LoS}}}{\bf{g}}_{e,\rm{NLoS}}^H{{\bf{h}}_{u,\rm{NLoS}}}}\right\}
    \\
    &={ {{\kappa^2 {\Gamma^2_e}}/{{M\left(\kappa  + 1\right)^2}}} } + {2\kappa }/{{{{\left( {\kappa  + 1} \right)}^2}}} + {1}/{{{{\left( {\kappa  + 1} \right)}^2}}}. \label{fk}
\end{aligned}
\end{equation}
Let $\mathbf{h}_{u,\rm{LoS}}^H{{\mathbf{g}}_{e,\rm{LoS}}}{\mathbf{g}}_{e,\rm{LoS}}^H{{\mathbf{h}}_{u,\rm{LoS}}}=\Gamma^2_e$, which is determined by the locations of $u$-th UAV-UE and $e$-th UAV-EVE. 
Additionally, other omitted terms of (\ref{fk}) are zeros.
Secondly, we can write $\mathbf{g}_e$ as $\mathbf{g}_e=\mathbf{h}_u+\tilde{\mathbf{h}}_e$ and we have 
\begin{equation}
    \tilde{\mathbf{h}}_e=\sqrt{\frac{\kappa}{\kappa+1}} \tilde{\mathbf{h}}_{u,\rm{LoS}}+\sqrt{\frac{1}{\kappa+1}}\tilde{\mathbf{h}}_{u,\rm{NLoS}},
\end{equation}
where $\tilde{\mathbf{h}}_{u,\rm{LoS}}=\mathbf{g}_{e,\rm{LoS}}-\mathbf{h}_{u,\rm{LoS}}$ is constant and $\tilde{\mathbf{h}}_{u,\rm{NLoS}}=\mathbf{g}_{e,\rm{NLoS}}-\mathbf{h}_{u,\rm{NLoS}}\sim\mathcal{CN}\left(\mathbf{0},2\mathbf{I}\right)$. 
Thus, we have 
\begin{equation}
\begin{aligned}
    &\mathbb{E}\left\{{\mathbf v_i^H{{\bf{g}}_e}{\bf{g}}_e^H{\mathbf v_i}}\right\}=\mathbb{E}\left\{{\mathbf v_i^H\left(\mathbf h_u-\tilde{\mathbf{h}}\right)\left(\mathbf h_u-\tilde{\mathbf{h}}\right)^H{\mathbf v_i}}\right\}\\
    &=\mathbb{E}\left\{{\mathbf v_i^H{{\tilde{\bf{h}}}_{\rm{LoS}}}{\tilde{\bf{h}}}_{\rm{LoS}}^H{\mathbf v_i}}\right\}+\mathbb{E}\left\{{\mathbf v_i^H{{\tilde{\bf{h}}_{\rm{NLoS}}}}{\tilde{\bf{h}}_{\rm{NLoS}}}^H{\mathbf v_i}}\right\}\\
    &=\frac{1}{N_{\rm{AN}}M}\left(\frac{\kappa}{\kappa+1}\left(2M-2\Gamma_e\right)+\frac{2}{\kappa+1}\right)=\frac{g\left(\kappa\right)}{N_{\rm{AN}}}.
\end{aligned}
		\label{var_an}
\end{equation}

Based on (\ref{fk}) and (\ref{var_an}), the SINR of message at $e$-th UAV-EVE can be calculated by
\begin{equation}
\text{SINR}_{e,\rm{m}}^{'}=\frac{\phi_{\rm s}^2\rho_{\rm m}^2\beta_u^{-1}\tilde{\beta}f\left(\kappa\right)}{\phi_{\rm s}^2\rho_{\rm t}^2\beta_u^{-1}\tilde{\beta}f\left(\kappa\right)+\phi_{\rm n}^2g\left(\kappa\right)+{\sigma_{\rm n}^2}/{P_{\rm T}\alpha_e}}. \label{SINR_e,m}
\end{equation}
Correspondingly, the SINR of tag at $e$-th UAV-EVE can be expressed as 
\begin{equation}
	\begin{aligned}
		\text{SINR}_{e,\rm{t}}^{'}&=\mathbb{E}\left\{ {\frac{{\phi_{\rm s}^2\rho_{\rm t}^2{\bf{w}}_u^H{{\bf{g}}_e}{\bf{g}}_e^H{{\bf{w}}_u}}}{{ \phi_{\rm n}^2\sum\limits_{i = 1}^{{N_{\rm{AN}}}} {\mathbf v_i^H{{\bf{g}}_e}{\bf{g}}_e^H{\mathbf v_i}}  + {\sigma_{\rm n}^2}/{P_{\rm T}\alpha_e}}}} \right\}\\&=\frac{\phi_{\rm s}^2\rho_{\rm t}^2\beta_u^{-1}\tilde{\beta}f\left(\kappa\right)}{\phi_{\rm n}^2g\left(\kappa\right)+{\sigma_{\rm n}^2}/{P_{\rm T}\alpha_e}}.
	\end{aligned}
\end{equation}
Similar to (\ref{P_s_u}) and (\ref{P_t_u}), the SERs of message and tag at $e$-th UAV-EVE can be demonstrated as 
\begin{equation}
	\begin{aligned}
		&{P_{e,\rm{m}}^{'}} =\\ &1- \frac{1}{2}{\left( {Q\left( {\frac{{\frac{\rho_{\rm m}}{\sqrt 2 } + {\rho_{\rm t}}}}{{ - {\sigma _{e}}}}} \right) + Q\left( {\frac{{\frac{{{\rho_{\rm m}}}}{{\sqrt 2 }} - {\rho_{\rm t}}}}{{ - {\sigma _{e}}}}} \right)} \right)Q\left( {\frac{{\frac{{{\rho_{\rm m}}}}{{\sqrt 2 }}}}{{ - {\sigma _{e}}}}} \right)},
	\end{aligned}
	\label{P_s_e}
\end{equation}
and 
\begin{equation}
	\begin{aligned}
		&{P_{e,\rm{t}}^{'}} = \\
		&\frac{1}{2} \left({\begin{aligned}
				{Q\left( {\frac{{{\rho_{\rm t}}}}{\sigma _{e} }} \right) - Q\left( {\frac{{\frac{\rho_{\rm m}}{\sqrt 2}  + {\rho_{\rm t}}}}{\sigma _{e} }} \right) + Q\left( {\frac{{\sqrt 2 {\rho_{\rm m}} + {\rho_{\rm t}}}}{\sigma _{e} }} \right)}\\
				{ +Q\left( {\frac{{\frac{\rho_{\rm m}}{\sqrt 2}  - {\rho_{\rm t}}}}{\sigma _{e} }} \right) - Q\left( {\frac{{\sqrt 2 {\rho_{\rm m}} - {\rho_{\rm t}}}}{\sigma _{e} }} \right)+Q\left( {\frac{{{\rho_{\rm t}}}}{\sigma _{e} }} \right) }
		\end{aligned}} \right),
	\end{aligned}
	\label{P_t_e}
\end{equation}
where 
\begin{equation}
	\sigma_{e}=\sqrt{\frac{\sigma_{\rm n}^2/P_{\rm T}\alpha_e+\phi_{\rm n}^2g\left(\kappa\right)}{2\phi_{\rm s}^2\beta_u^{-1}\tilde{\beta}f\left(\kappa\right)}}.
\end{equation}

\subsection{Security optimization}
The classical secrecy evaluation is the ergodic sum secrecy rate, expressed as 
\begin{equation}
	R=\sum_{k=1}^{K}\left(\log_2\left(1+\text{SINR}_{k,\rm{m}}\right)-\log_2\left(1+\text{SINR}^{'}_{k,\rm{m}}\right)\right)^{+}.\label{R}
\end{equation}
However, this evaluation $R$ is not suitable for assessing the secrecy performance of our scheme, since it only considers the secrecy of the ciphertext message but does not consider the effects of authentication and encoding. 
Therefore, we propose our modified ergodic sum secrecy rate to evaluate the specific secrecy performance of our proposed TBE scheme. 
At UAV-UEs, only authenticated message blocks will be accepted and correctly decoded. 
Thus, the user data rate should be weighted with the authentication probability, which travels from $0$ to $1$ with the decrease of $P_{u,\rm{t}}$. 
At UAV-EVEs, plaintext is obtained by decoding the ciphertext according to the detected tag. 
Each tag bit contains one bit of decoding information, and there are $\log_2\left(1/\left(1 - P_{e,\rm{t}}^{'}\right)\right)$ information remaining after transmission. 
So, the wiretap information amount ratio is expressed as 
\begin{equation}
	I_e=1-\log_2\left(1/\left(1-P_{e,\rm{t}}^{'}\right)\right)=1+\log_2\left(1-P_{e,\rm{t}}^{'}\right).
\end{equation}
Obviously, $I_e$ has the same property that travels monotonically from $0$ to $1$ with the decrease of $P_{e,\rm{t}}^{'}$. 
Under the effect of authentication and encoding, the ergodic sum secrecy rate of the plaintext can be modified as 
\begin{equation}
	\begin{aligned}
			&R_{\rm{sec}}=\\
			&\sum_{k=1}^{K}\left(P_{k,\rm{d}}\log_2\left(1+\text{SINR}_{k,\rm{m}}\right)-I_k\log_2\left(1+\text{SINR}^{'}_{k,\rm{m}}\right)\right)^{+}.
	\end{aligned}\label{R_s}
\end{equation}
Such modified secrecy rate can be referred from \cite{PD1,PD2,Shannon_secrecy}.
The data rate is weighted by detection probability in cognitive radio systems\cite{PD1,PD2}.
The wiretap rate weighted by wiretapped tag's information amount ratio is derived based on the classical one-time-pad secrecy proposed by Shannon \cite{Shannon_secrecy}. 

To verify the security enhancement of our proposed scheme and find suitable power allocations, we formulate an unconstrained ergodic sum secrecy rate $R_{\rm{sec}}$ maximization problem. 
Let $\rho_{\rm m}^2=\rho$ ($\rho_{\rm t}^2=1-\rho$) and $\phi_{\rm s}^2=\phi$ ($\phi_{\rm n}^2=1-\phi$). 
The optimization problem is addressed as 
\begin{equation}
	\begin{array}{*{20}{c}}
		{\mathop {\arg \max }\limits_{\mathbf{p}=\left[\rho, \phi\right]^T } }&R_{\rm{sec}}
		.
	\end{array}\label{R_s1}
\end{equation}
In practice, receptions at UAV-UEs have been balanced by precoding in (\ref{precoder}) and their wiretap risks are comparable ($\vartheta_e-\theta_u|_{u=e}=\Delta\theta, \forall e$). 
The secrecy performance of each UAV-UE and UAV-EVE pair varies synchronously and their achievable optimal secrecy rate are always non-negative, thus we removes $\left(\cdot\right)^{+}$ \cite{secrecy_rate}. 
Let's denote sum data rate $R_{\rm{U}}$ and sum wiretap rate $R_{\rm{E}}$ by 
\begin{equation}
	R_{\rm U}=\sum_{k=1}^{K}\log_2\left(1+\text{SINR}_{k,\rm{m}}\right) \cdot P_{k,\rm{d}}
	,\label{Rk}
\end{equation}
and 
\begin{equation}
	R_{\rm E}=\sum_{k=1}^{K}\log_2\left(1+\text{SINR}^{'}_{k,m}\right) \cdot I_k
	.
\end{equation}
Then, the problem (\ref{R_s1}) can be transformed as 
\begin{equation}
	\begin{array}{*{20}{c}}
		{\mathop {\arg \min }\limits_{\mathbf{p} } }&{R_{\rm{E}}\left(\mathbf p\right)-R_{\rm{U}}\left(\mathbf p\right)
			=-R_{\rm{U}}\left(\mathbf p\right)-\left(-R_{\rm{E}}\left(\mathbf p\right)\right)
			}.
	\end{array}\label{R_s2}
\end{equation}
Specifically, this is a classic difference-of-convex (DC) programming, so the DC algorithm (DCA) can be used to find suitable power allocations \cite{DCA}. 
The quasi-concavity proofs of $R_{\rm{U}}$ and $R_{\rm{E}}$ are given in Appendix A and B, respectively, and $-R_{\rm{U}}$ and $-R_{\rm{E}}$ are quasi-convex. 
The main procedure of DCA is to transform the problem into its convex dual form, shown as 
\begin{equation}
	\begin{array}{*{20}{c}}
		{\mathop {\arg \min }\limits_{\mathbf q=\left[\varrho, \varphi\right]^T } }&{R_{\rm{U}}^{*}\left(\mathbf q\right)-R_{\rm{E}}^{*}\left(\mathbf q\right)}
		,\label{dual}
	\end{array}
\end{equation}
where $\mathbf q=\left[\varrho, \varphi\right]^T$ is a temporary variable vector, and $R_{\rm{U}}^{*}$ and $R_{\rm{E}}^{*}$ are conjugate functions of $R_{\rm{U}}$ and $R_{\rm{E}}$. 
The conjugate transform is expressed as 
\begin{equation}
R_{\rm{U}(\rm{E})}^{*}\left(\mathbf q\right)=\sup\left\{<\mathbf{p},\mathbf{q}>-R_{\rm{U}(\rm{E})}\left(\mathbf p\right)\right\}.
\end{equation}
So far, (\ref{dual}) is a convex problem and (\ref{R_s1}) can be solved by an iterative DCA, given in \textbf{Algorithm} \ref{alg:alg1}. 
\begin{algorithm}[t]
	\caption{Iterative DCA for Finding the Optimal $\mathbf p$.}\label{alg:alg1}
	\begin{algorithmic}
		\STATE \textbf{Initialize:} $\Omega=0$, $\mathbf p^{\Omega}=\left[1,1\right]^T$, $\Omega_{\max}$, $\epsilon$
		\STATE \textbf{while} 
		\STATE  \hspace{0.1cm} $\mathbf{q}^{\Omega}=-\nabla R_{\rm{E}}\left(\mathbf p^{\Omega}\right)$
		\STATE  \hspace{0.1cm} $\mathbf p^{\Omega+1}=\mathop {\arg \min }\limits_{\mathbf p }~R_{\rm{E}}\left(\mathbf p^{\Omega}\right)-R_{\rm{U}}\left(\mathbf p\right)-<\mathbf p-\mathbf p^{\Omega},\mathbf q^{\Omega}>$
		\STATE  \hspace{0.1cm} $\Omega=\Omega+1$
		\STATE \hspace{0.1cm} \textbf{if} $\Omega\ge\Omega_{\max}~\text{or}~ \|\mathbf p^{\Omega}-\mathbf{p}^{\Omega-1}\|\le\epsilon$
		\STATE \hspace{0.2cm} \textbf{break}
		\STATE \hspace{0.1cm} \textbf{end}
		\STATE \hspace{0cm} \textbf{end}
		\STATE $\mathbf p_{\text{opt}}=\mathbf p^{\Omega}$
	\end{algorithmic}
\end{algorithm}

Moreover, to demonstrate our proposed TBE scheme able to enhance security from the perspective of reliability, the security problem is formulated as a transmission efficiency optimization problem with the constraint of secrecy and authentication. 
The transmission efficiency is reflected as the authentication (transmission) fail probability (AFP), which is given as 
\begin{equation}
	\text{AFP}_u=1-\left(1-\text{BLER}_u\right)\cdot P_{u,\rm{d}}, 
\end{equation}
where $\text{BLER}_u=1-\left(1-P_{u,\rm{m}}\right)^T$ is the block error ratio of message \cite{AFP}. 
In practice, $\text{AFP}=\frac{1}{K}\sum_{k = 1}^{{K}}\text{AFP}_k$ is chosen as the objective. 
The authentication constraint is the average of authentication probability $P_{\rm{d}}=\frac{1}{K}\sum_{k = 1}^{{K}}P_{k,\rm{d}}$. 
Due to the reception balance for UAV-UEs, we have $\text{AFP}=\text{AFP}_u$ and $P_{\rm{d}}=P_{u,\rm{d}},\forall u$. 
The secrecy constraint is built as the wiretap  probability as well as the average SER of the ciphertext at UAV-EVEs which is defined as 
\begin{equation}
		P_{\rm{w}}=1-\frac{1}{K}\sum\limits_{k = 1}^{{K}}\left(1-P_{k,\rm{m}}^{'}\right)\left(1-P_{k,\rm{t}}^{'}\right).
\end{equation}

The AFP minimization problem is formulated as 
\begin{equation}
	\begin{array}{*{20}{c}}
		{\mathop {\arg \min }\limits_{\mathbf{p}} }&{{\text{AFP}}}
		\\
		{\rm{s.t.}}&{  {P_{\rm w}} \ge   {P_{\rm{w},0 }}},
		\\
		{}&{  {P_{\rm{d}}} 
			\ge   {P_{\rm{d},0 }}}
		,
	\end{array}\label{constrained AFP}
\end{equation}
where $P_{\rm{w},0}$ and $P_{\rm{d},0}$ are thresholds of wiretap and authentication constraints. 

With the decrease of $\phi$, injected AN increases. 
AFP and $P_{\rm w}$ increase while $P_{\rm{d}}$ decreases monotonously. 
This trend indicates the concession of transmission and authentication for less information leakage. 
With the decrease of $\rho$, more power is allocated to the tag. 
AFP decreases initially and then increases. 
In addition, $P_{\rm{d}}$ increases monotonously, and $P_{\rm{w}}$ decreases monotonously. 
Above the varying trend with $\rho$ and $\phi$, AFP is quasi-convex, while $P_{\rm{w}}$ and $P_{\rm{d}}$ are affine. 
The quasi-convexity of AFP is proved in Appendix C.

This problem can be reformulated as an unconstrained optimization problem using Lagrange method $\left(\mathbf{\lambda} = \left[\lambda_1, \lambda_2\right]^T\right)$, expressed as
\begin{equation}
	\begin{aligned}
		L\left(\mathbf p,\mathbf{\lambda}\right)=\text{AFP}-\lambda_1\left(P_{\rm{w},0}-P_{\rm{w}}\right)-\lambda_2\left(P_{\rm{d},0}-P_{\rm{d}}\right).
	\end{aligned}\label{lag}
\end{equation}
For solving this problem, the Karush–Kuhn–Tucker (KKT) conditions are shown as
\begin{subequations}
			\begin{align}
			&	\frac{{\nabla }}{{\nabla \mathbf{p}}}\text{AFP} + {\lambda _1}\frac{\nabla }{{\nabla \mathbf{p}}}\left(P_{\rm{w},0}-P_{\rm{w}}\right) + {\lambda _2}\frac{\nabla }{{\nabla \mathbf{p}}}\left(P_{\rm{d},0}-P_{\rm{d}}\right) = \mathbf{0},\label{e1}\\
			&{\lambda _1} \ge 0,\\
			&{\lambda _2} \ge 0,\\
			&P_{\rm{w},0}-P_{\rm{w}}\le 0,\\
			&P_{\rm{d},0}-P_{\rm{d}}\le 0,\\
			&\lambda_1\left(P_{\rm{w},0}-P_{\rm{w}}\right)=0,\\
			&\lambda_2\left(P_{\rm{d},0}-P_{u,\rm{d}}\right)=0.
		\end{align}
\end{subequations}

However, AFP is quasi-convex, which means (\ref{e1}) can not be solved due to the existence of saddle points. 
To overcome this challenge, we convert this two-variable ($\rho,\phi$) problem into a one-variable $\rho$ problem in some complementary slackness constraint cases, and the one-variable problems are solved by the bisection searching method or gradient descent algorithm.  
According to complementary slackness, the solution process can be simplified and divided into four cases: 
\begin{itemize}
    \item[1)]
    In the first case, both constraints are satisfied when the optimal value for the unconstrained objective is achieved. 
Here, $\lambda_1 = 0$ and $\lambda_2 = 0$. 
The optimal $\phi^{*} = 1$ and the optimal $\rho^{*}$ can be obtained by the one-dimension iterative bisection algorithm given in \textbf{Algorithm} \ref{alg3}. 
\item[2)]
In the second case, the authentication constraint is satisfied, but the secrecy constraint is not when the optimal value for the unconstrained objective is achieved. 
In this scenario, $\lambda_1 > 0$ and $\lambda_2 = 0$. 
The optimal solution is achieved when $P_{\rm{w}}=P_{\rm{w},0}$, and we have $\phi= P_{\rm{w}}^{-1}\left(P_{\rm{w},0}|\rho\right)$, where $P_{\rm{w}}^{-1}$ indicates the inverse function of $P_{\rm{w}}$. 
Then, the optimal $\rho^{*}$ can be obtained by \textbf{Algorithm} \ref{alg3} and $\phi^{*}=P_{\rm{w}}^{-1}\left(P_{\rm{w},0}|\rho^{*}\right)$.

\item[3)]
In the third case, the secrecy constraint is satisfied, but the authentication constraint is not when the optimal value for the unconstrained objective is achieved. 
Here, $\lambda_1 = 0$ and $\lambda_2 > 0$. 
The optimal solution is achieved when $P_{\rm{d}}=P_{\rm{d},0}$, and we have $\phi= P_{\rm{d}}^{-1}\left(P_{\rm{d},0}|\rho\right)$, where $P_{\rm{d}}^{-1}$ indicates the inverse function of $P_{\rm{d}}$. 
Then, the optimal $\rho^{*}$ can be obtained by \textbf{Algorithm} \ref{alg3} and $\phi^{*}=P_{\rm{d}}^{-1}\left(P_{\rm{d},0}|\rho^{*}\right)$. 

\item[4)]
In the fourth case, neither constraint is satisfied when the optimal value for the unconstrained objective is achieved. 
In this situation, $\lambda_1 > 0$ and $\lambda_2 > 0$. 
The optimal $\left(\rho^{*}, \phi^{*}\right)$ can be obtained by solving
\begin{equation}
	\left\{ \begin{array}{l}
		P_{\rm{w},0}-P_{\rm{w}}=0,\\
		P_{\rm{d},0}-P_{\rm{d}}=0.
	\end{array} \right. 
\end{equation}
\end{itemize}

\begin{algorithm}[t]
	\caption{Iterative bisection algorithm for optimal $\rho^{*}$. }\label{alg3}
	\begin{algorithmic}
		\STATE \textbf{Initialize:} $\Omega$=0, $\rho_{\max}^{\Omega}=1$, $\rho_{\min}^{\Omega}=0.9$, $\epsilon$
		\STATE \textbf{while} ($\rho_{\max}^{\Omega}-\rho_{\min}^{\Omega}>\epsilon$)
		\STATE \hspace{0.1cm} rand $\rho_1^{\Omega}$, $\rho_2^{\Omega}$ ($\rho_{\min}^{\Omega}<\rho_1^{\Omega}<\rho_2^{\Omega}<\rho_{\max}^{\Omega}$)
		\STATE \hspace{0.1cm} \textbf{if} $\text{AFP}\left(\rho_1^{\Omega}\right)<\text{AFP}\left(\rho_2^{\Omega}\right)$
		\STATE \hspace{0.2cm} $\rho_{\min}^{\Omega+1}=\rho_{\rm{min}}^{\Omega}$, $\rho_{\max}^{\Omega+1}=\rho_{2}^{\Omega}$
		\STATE \hspace{0.1cm} \textbf{else}
		\STATE \hspace{0.2cm} $\rho_{\min}^{\Omega+1}=\rho_{1}^{\Omega}$, $\rho_{\max}^{\Omega+1}=\rho_{\max}^{\Omega}$
		\STATE \hspace{0.1cm} \textbf{end}
		\STATE \hspace{0.1cm} $\Omega=\Omega+1$
		\STATE \textbf{end}
		\STATE
		$\rho^{*}=\left(\rho_{\max}^{\Omega}-\rho_{\min}^{\Omega}\right)/2$
	\end{algorithmic}
\end{algorithm}

\begin{table}[!b]
	\caption{Simulation parameter setting.\label{setting}}
	{\begin{tabular*}{20pc}{@{\extracolsep{\fill}}cc@{}}\hline
			Parameter  &Value \\
			\hline\hline
			Transmit power ($P_{\rm T}$)&5 dBm      \\
			Central frequency ($f_{\rm c}$)& 2.4 GHz   \\
			Relative bandwidth (BW)&  300 MHz   \\
			Thermal noise ($N_0$)& -174 dBm/Hz   \\
			Noise Figure (NF) & 9 dB \\
			Number of antennas at BS ($M$)& 64   \\
			Height of UAV-UEs ($h_{\rm UE}$) & 100 m\\
			Height of UAV-EVEs ($h_{\rm EVE}$) & 60 m / 80 m\\
			Number of UAV-UEs ($K$)& 4\\
			Horizon distance range of UAV-UEs ($l_u$) & [10 m, 100 m]\\
			Time Slot ($T$)& 160   \\
			Bit of the key & 64  \\
			$\kappa$-factor ($\kappa$)& 30 dB  \\
			False alarm probability ($P_{\rm f}$)&0.001\\
			\hline\hline
	\end{tabular*}}{}
\end{table}
\section{Numerical simulation}\label{simulation}
In this section, the numerical simulation results with ``SIM" are performed by MATLAB, and theoretical analyses with ``THR" are also given. 
The main communication parameter settings are provided in Table \ref{setting}. 
The legitimate transmission is affected by transmission parameters and UAV-UEs' number and locations, which have been fixed and shown in the table. 
The UAV-EVEs are deployed at several locations that provide advantages for wiretapping. 
The legitimate and illegitimate transmissions are investigated first, and then, the two optimization problems are studied.

\begin{figure}[!t]\centering
	\includegraphics[width=3.0in]{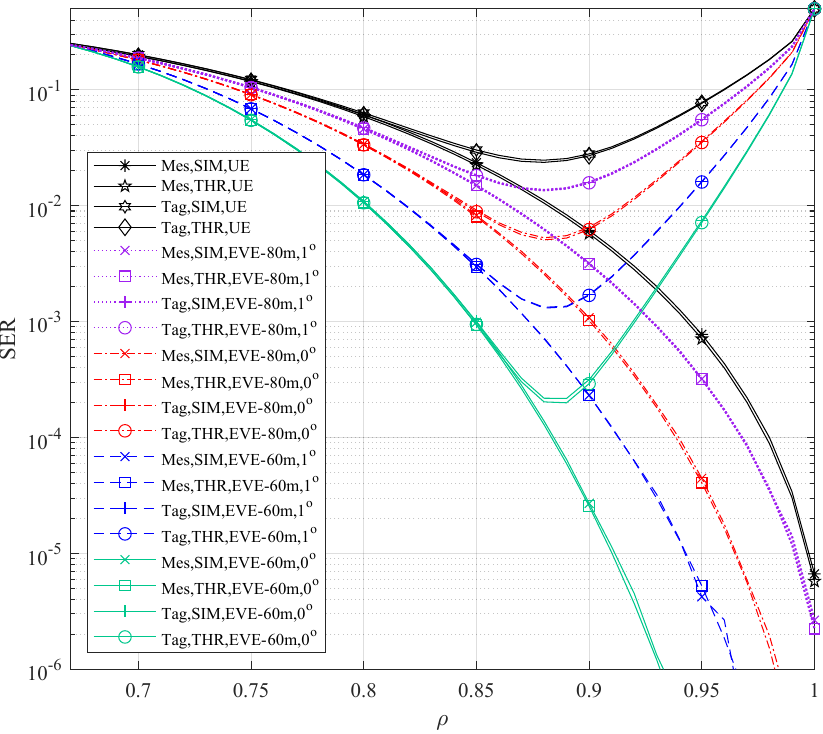}
	\caption{Average SERs of message and tag at UAV-UEs and UAV-EVEs versus $\rho$ when AN is not injected ( e.g., the curve ``Mes,SIM,EVE-{\text{80m},$1^{\circ}$}'' indicates the simulation of messages' average SER at UAV-EVEs with $h_{\rm{EVE}}=\text{80 m}$ and $\Delta\theta=1^{\circ}$. Specifically, ``Mes/Tag'' indicates message/tag, ``SIM/THR'' indicates simulation/theory, ``EVE/UE'' indicates eavesdropper/user, ``80 m/60 m'' is the value of $h_{\rm{EVE}}$, and $0^{\circ}/1^{\circ}$ is the value of $\Delta\theta$). 
	}\label{SER_rho}
\end{figure}

In Fig. \ref{SER_rho}, SERs of message and tag at UAV-UEs and UAV-EVEs are presented, where AN has not been injected yet. 
The presented SERs are average performance of $K$ UAV-UEs and that of $K$ UAV-EVEs. 
The average SER of all UAV-UEs is same to SER of each UAV-UE due to the transmission balance, while there are slight differences between the SERs of different eavesdroppers. 
The message transmission is the foundation of communication, while the tag transmission provides guarantee for message security. 
There are four wiretap cases where UAV-EVEs are deployed at an altitude of $\text{80 m}$ or $\text{60 m}$ with $\Delta\theta$ equaling $0^{\circ}$ or $1^{\circ}$, which is lower than $h_{\rm{UE}}=\text{100 m}$. 
In these four cases, it is clear that both BERs of message and tag at UAV-EVEs are lower (better) than those at UAV-UEs, i.e., the legitimate transmission is under severe threat of eavesdropping. 
The varying trends of SERs at UAV-UEs and UAV-EVEs to $\rho$ are the same. 
At $\rho=$ 1, no power is allocated to the tag, resulting in a tag SER of $0.5$ and the lowest SER for the ciphertext. 
As $\rho$ decreases, the power allocated to the tag increases while that to the message decreases. 
The reception of message increases monotonically resulting from the power loss and the tag's interference. 
The tag's SER initially decreases due to the benefit of more power when the reception of message is much more reliable than the tag. 
As $\rho$ decreases significantly, the degradation of message reception begins to negatively impact message reception, which in turn affects tag recovery. 
When $\rho$ is small enough, both message's and tag's SERs increase due to interference with each other. 
It is also noteworthy that the theoretical analyses and simulation results show a high degree of consistency. 
Communications' performance is related to both message's and tag's transmissions, so the power allocations should be well-optimized. 

\begin{figure}[!t]\centering
	\includegraphics[width=3.0in]{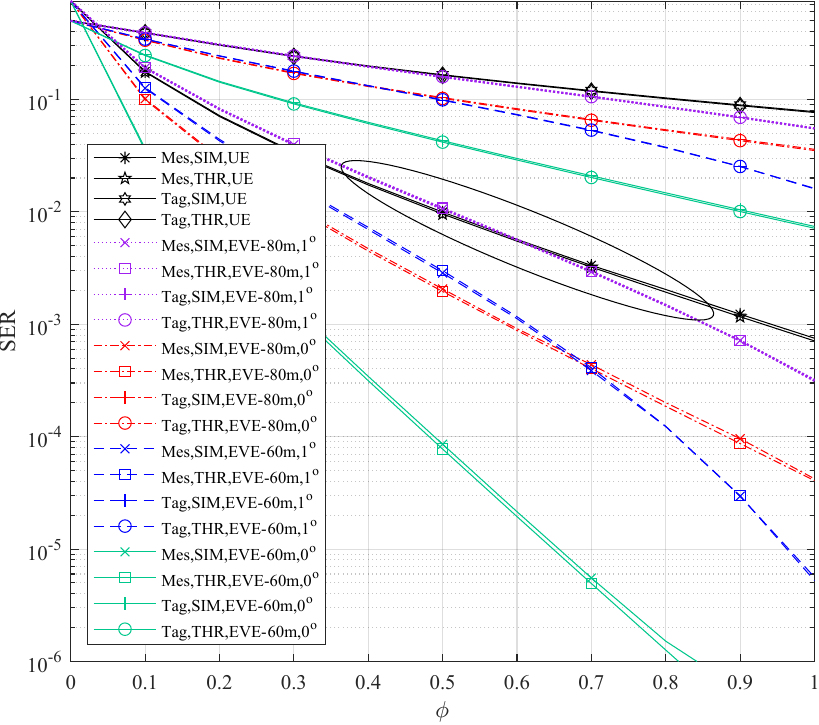}
	\caption{Average SERs of message and tag at UAV-UEs and UAV-EVEs versus $\phi$ with $\rho=0.95$. 
		Low-risk scenario is with $h_{\rm{EVE}}=80$ m and $\Delta\theta=1^{\circ}$, where AN injection works (able to degrade EVE's SER to be worse than user's SER as circled). Other scenarios are high-risk scenarios where AN injection fails.  }\label{SER_phi}
\end{figure}

\begin{figure}[!t]\centering
	\includegraphics[width=3.0in]{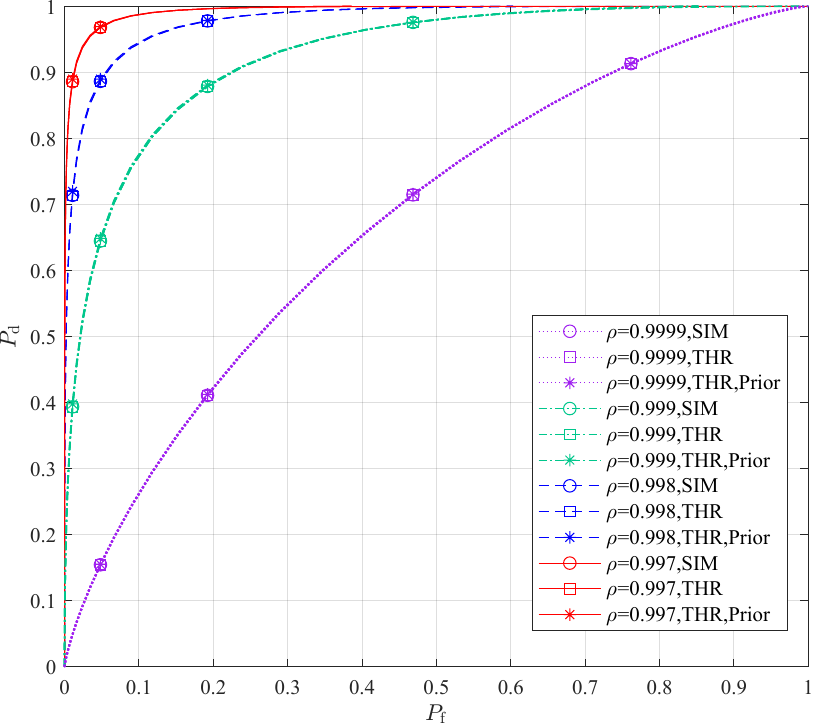}
	\caption{ROC at different $\rho$ with $\phi=1$ (``THR,Prior'' indicates the prior $P_{\rm f}$ expression given in \cite{SIMO}, which considers $P_{u,\rm b}$ in (\ref{pf}) as $0$). 
		Authentication performance increases as more power is allocated to the tag. Every three lines (``SIM'', ``THR'', and ``THR,Prior'') with the same $\rho$ are consistent. }\label{ROC}
\end{figure}

\begin{figure*}[!t]\centering
	\includegraphics[width=6in]{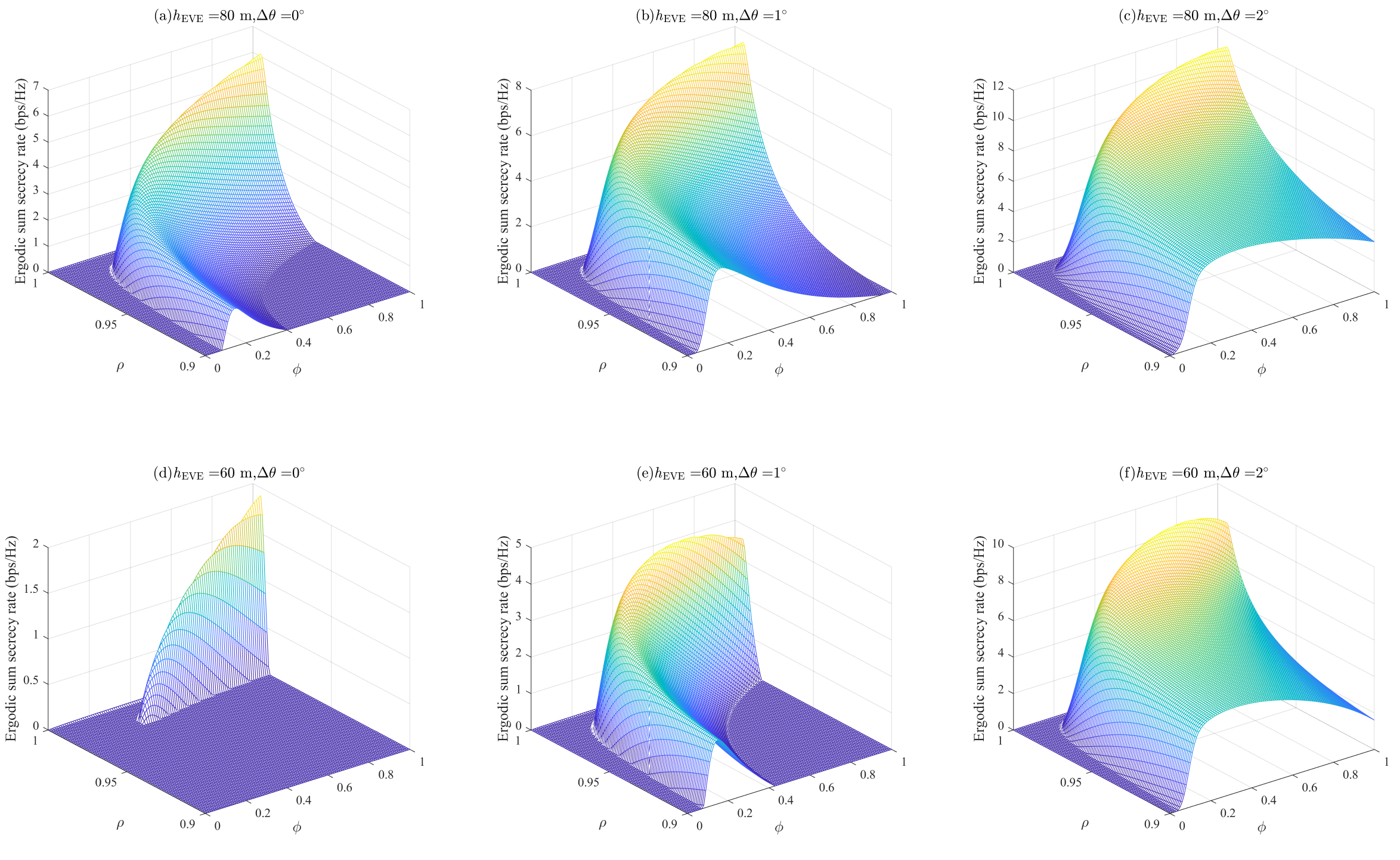}
	\caption{Ergodic sum secrecy rate varying with $\rho$ and $\phi$ at different wiretap scenarios, where the threshold of $P_{\rm f}$ is set to 0.001. 
	The special characteristic of DC programming can be observed as the one or two stationary points from ($0.9,1$) varying towards ($1,1$). }\label{R_6}
\end{figure*} 

In Fig. \ref{SER_phi}, SERs varying with $\phi$ are investigated, where $\rho$ is set to 0.95. 
First of all, at $\phi=$ 1, there is no AN injected and all UAV-EVEs' SERs are lower than UAV-UEs' SERs. 
With the decrease of $\phi$, all SERs get worse. 
Observed from the specially circled comparison, with AN injection, the SER of message at UAV-UE becomes lower than that at UAV-EVEs when UAV-EVEs are deployed with $h_{\rm{EVE}}=\text{80 m}$ and $\Delta\theta=1^{\circ}$. 
This phenomenon reveals the traditional effect of AN on secrecy in a scenario with low risk. 
In contrast, other three scenarios are defined as high-risk scenarios, where AN injection can not degrade UAV-EVEs' message SERs to be lower than UAV-UEs', and secrecy is not achievable by AN injection. 
Facing these scenarios with high risk, our proposed TBE scheme makes secrecy possible. 
In our scheme, secrecy is not achieved by introducing AN to degrade the message reception directly but by degrading the tag wiretap to reduce information leakage through wiretapped tag-based decoding. 
The simulation and theoretical results are consistent. 
Because SERs are related to SINRs, the consistency of SERs in Fig. \ref{SER_rho} and Fig. \ref{SER_phi} also verify that numerical and theoretical SINRs are consistent. 
These two evaluations are the foundations of overall performance analysis, and their varying trends with $\rho$ and $\phi$ provide the theoretical basis for our optimization problems.

In Fig. \ref{ROC}, ROC varying with $\rho$ is illustrated to present the authentication performance. 
There are two indices $P_{\rm f}$ ($P_{\rm{f}}=\sum_{k=1}^{K}P_{k,\rm{f}}$ and $P_{\rm{f}}=P_{u,\rm{f}},\forall u$) and $P_{\rm{d}}$, associated by the inherent variable $\eta$ according to (\ref{pd}) and (\ref{pf}). 
Specifically, $P_{\rm f}$ indicates the ability to reject false messages, while $P_{\rm d}$ indicates the accuracy of authenticating legitimate messages. 
Reliable authentication is realized by choosing suitable threshold $\eta$ and tag power allocation for a high $P_{\rm{d}}$ and a small $P_{\rm{f}}$. 
Generally, $P_{\rm f}$ is a small value and it directly determines the value of $\eta$ according to (\ref{pf}). 
And then, all messages with errors less than $\eta$ will be accepted. 
The higher the achievable $P_{\rm{d}}$ at the threshold $\eta$, the better the authentication performance. 
Observing from this figure, with $\rho$ traveling from 0.9999 to 0.997, the curve shifts from diagonal to the upper left. 
Fundamentally, this improvement is driven by the decrease in the SER of tag, which decreases monotonically when the tag is covert. 
The highest authentication probability can be achieved at the stationary point of tag's SER, but this results in a significant deterioration of message transmission. 
Additionally, the simulation and theoretical results are consistent. 
Furthermore, the prior $P_{\rm f}$ expression given \cite{SIMO} is also presented, and there is almost no difference compared to our simulation and numerical results. 
This consistency proves that $P_{u,{\rm b}}$ is very small and our proposed new tag generation has a tiny degradation in authentication. 
In summary, our proposed TBE scheme has no impact on authentication, and power allocations should not only pursue authentication accuracy but also consider the overall communication performance and secrecy level. 

The secrecy performance of our proposed scheme aided by secrecy and authentication mechanisms can be evaluated through the ergodic sum secrecy rate $R_{\rm{sec}}$ given in (\ref{R_s}). 
In Fig. \ref{R_6}, ergodic sum secrecy rates $R_{\rm{sec}}$, varying with $\rho$ and $\phi$, are investigated in six scenarios with $h_{\rm{EVE}}=\text{80 \rm{m}}$ or $h_{\rm{EVE}}=\text{60 \rm{m}}$ and $\Delta\theta=0^{\circ}$ or $\Delta\theta=1^{\circ}$ or $\Delta\theta=2^{\circ}$. 
On one hand, the varying trend provides guidance on searching suitable power allocations. 
On the other hand, a larger achievable maximum secrecy rate, compared to other schemes, serves as the strongest evidence supporting the effectiveness of our proposed security scheme.  
The threshold of $P_{\rm f}$ is set to 0.001 shown in Tab. \ref{setting}. 
From these figures, several important phenomena can be observed. 
Firstly, the ergodic sum secrecy rate maximization problem is not convex but is a DC programming problem, because there are one or two stationary points on the line between any two points in the domain. 
In each figure, the objective is quasi-convex with respect to both the $\rho$ and $\phi$ axes, respectively. 
However, it is not convex when considering both $\rho$ and $\phi$ increasing simultaneously, e.g., as $\rho$ and $\phi$ vary from (0.9, 0) towards (1, 1), the evaluation increases at first, then decreases, then increases, and then decreases. 
This two-stationary-points characteristic is a classic symbol of the DC programming problem. 
The maximum value of $R_{\rm{sec}}$ always exists and is unique, and the suitable power allocations can be obtained according to the DCA method given in Algorithm \ref{alg:alg1}. 
Secondly, with the decrease of $h_{\rm{EVE}}$ and $\Delta\theta$, the maximum value of $R_{\rm{sec}}$ decreases. 
It means that secrecy becomes difficult with the wiretap risk increasing resulting from $\text{SINR}_{e,\rm{m}}^{'}$ given in (\ref{SINR_e,m}) increasing. 
In Fig. \ref{R_6}(a)-(d), the maximum value of $R_{\rm{sec}}$ is achieved when $\rho$ is close to 1 and $\phi=$ 1, meaning that AN is not required. 
In Fig. \ref{R_6}(e)-(f), the maximum value of $R_{\rm{sec}}$ is achieved when $\phi\ne\text{1}$. 
It reveals that AN is required when the wiretap channel gain is larger than the legitimate channels' gain and there are significant differences between legitimate and illegitimate channels. 
Also, the  potential of having two peaks in DC programming should be mentioned. 
Thirdly, our proposed TBE can achieve a higher secrecy rate than AN injection. 
Specifically, in high-risk scenarios, where channels are similar and AN injection fails to achieve secrecy, TBE is effective to substitute AN injection for secrecy rate achieving. 
The fundamental of TBE is the different utilization of the received tag. 
At UAV-UEs, the received fuzzy tag serves the reference for authentication while the regenerated accurate tag is the reference for decoding. 
At UAV-EVEs, the wiretapped tag would be used for decoding directly, which is significantly less reliable than decoding at UAV-UEs. 
This novel mechanism does not rely on channel differences, allowing TBE to achieve higher secrecy than pure AN injection in high-risk scenarios. 
The figures above illustrate the achievable secrecy rate and the corresponding optimal power allocations.


\begin{figure}[!t]\centering
	\includegraphics[width=3.0in]{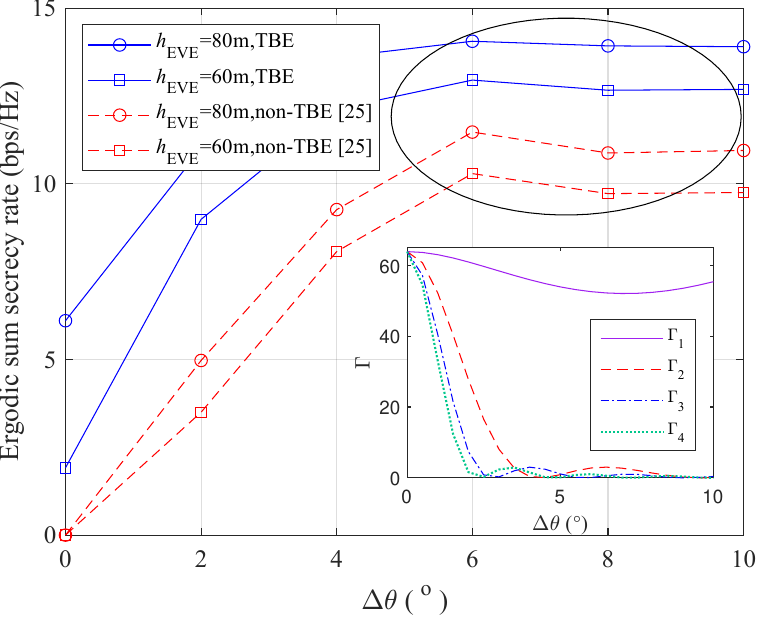}
	\caption{Comparison of ergodic sum secrecy rate of our proposed TBE with the prior AN-aided tag embedding scheme given in \cite{UAV}. The inserted picture is to explain that the swing occurring at large angle differences is caused by the swings of $\Gamma_e=\|\mathbf{h}_{u,\rm{LoS}}^H\mathbf{g}_{e,\rm{LoS}}\||_{u=e}$, which is important component of wiretap evaluations in Section \ref{wiretap analysis}. $\Gamma_e$ indicates the channel correlation determined by the initial angle and the varying angle.  
	}\label{com}
\end{figure}

In Fig. \ref{com}, the comparison of our proposed TBE scheme and the prior AN-aided tag embedding scheme is given. 
The reference method is the scenario provided in \cite{UAV} where the number of eavesdroppers is expanded to the same number of users. 
All basic techniques, including ZF precoder, null-space AN, and tag embedding, are the same among these two schemes. 
Besides, the communication scenario and deployments of UAV-UEs and UAV-EVEs are the same. 
The difference between them is whether the TBE scheme is used or not. 
All ergodic sum secrecy rates are maximized values after choosing optimized power allocations. 
The first superiority of TBE is that it improves the secrecy rates of all scenarios. 
The average values of the ergodic sum secrecy rates when $\Delta\theta\ge6^{o}$ (scenarios with low risks) are improved from 11.1 bps/Hz and 9.9 bps/Hz to 14.0 bps/Hz and 12.8 bps/Hz, respectively. 
The improvement ratios are 25.8$\%$ and 28.7$\%$, respectively. 
The second superiority is that the TBE scheme overcomes the limitation of no secrecy when $\Delta\theta=$ 0 and $h_{\rm EVE}\le h_{\rm UE}$ (high-risk scenarios). 
The ergodic sum secrecy rates are improved from 0 to 6.1 bps/Hz and 1.9 bps/Hz, respectively. 
Additionally, the inserted figure presents the varying of $\Gamma_e$ ($e=1,\cdots,K$), which is important component of $\text{SINRs}$ and SERs of message and tag at UAV-EVEs. 
The swing of $\Gamma_e$ is the main reason of the swing of ergodic sum secrecy rate when $\Delta\theta$ is large. 
From this comparison, it is clear that our proposed TBE scheme greatly enhances the secrecy performance of UAV communications.

\begin{figure}[!t]\centering
	\includegraphics[width=3.0in]{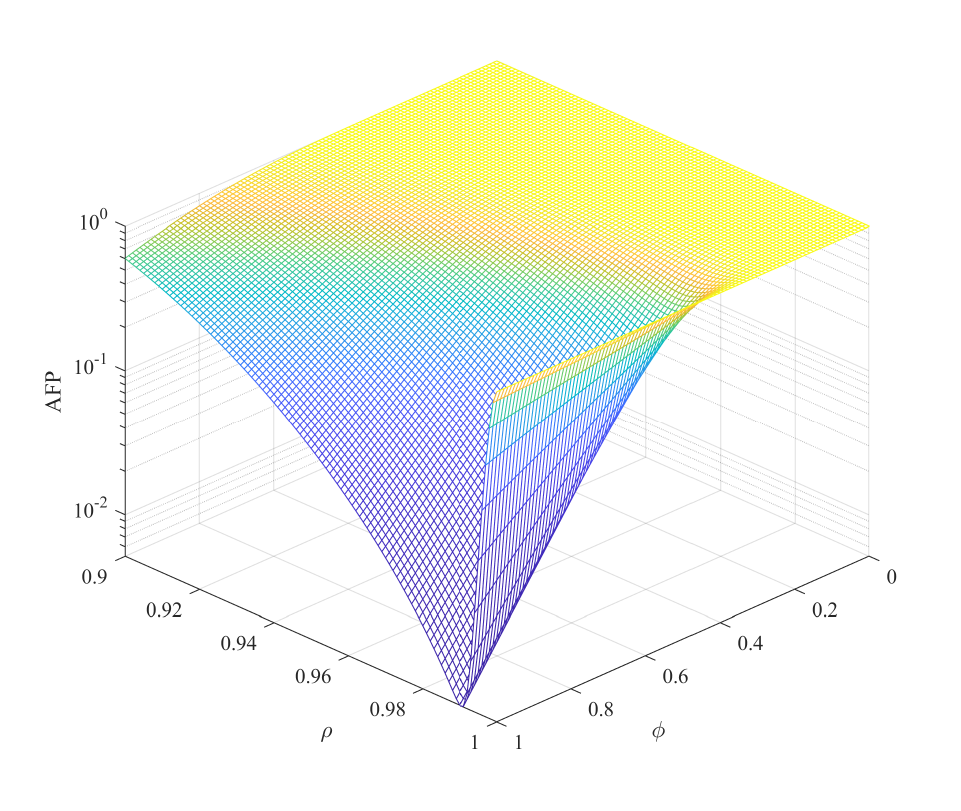}
	\caption{AFP versus $\rho$ and $\phi$, where the valley value indicates the optimal value without considering constraints. }\label{AFP}
\end{figure}

Analyzing from the view of transmission efficiency, the evaluation AFP varying with $\rho$ and $\phi$ is given in Fig. \ref{AFP}. 
The lower AFP is, the better the transmission efficiency is. 
Observing from the figure, AFP increases monotonically with the injection of AN while it is quasi-convex to $\rho$. 
This indicates that AFP is quasi-convex and there is a global minimum value. 
This valley represents the lower bound of transmission efficiency which can be achieved under loose secrecy and authentication constraints. 
Worthy to be mentioned that the global valley is always achieved when AN is not injected and the optimal $\rho^{*}$ is determined by the threshold of false alarm $\eta$. 
When global minimum AFP is achieved, $P_{\rm{d}}$ is 0.995 and $P_{\rm{w}}$ is related to the locations of UAV-EVEs. 
In addition, both authentication and secrecy constraints are affine. 
Specifically, $P_{\rm{d}}$ increases with $\phi$ while decreases with $\rho$, and $P_{\rm{w}}$ decreases with $\phi$ and $\rho$. 
The quasi-convexity of the objective and the affine properties of constraints provide theoretical foundations for solving the AFP minimization through the Lagrange method. 

\begin{figure}[!t]\centering
	\includegraphics[width=3.0in]{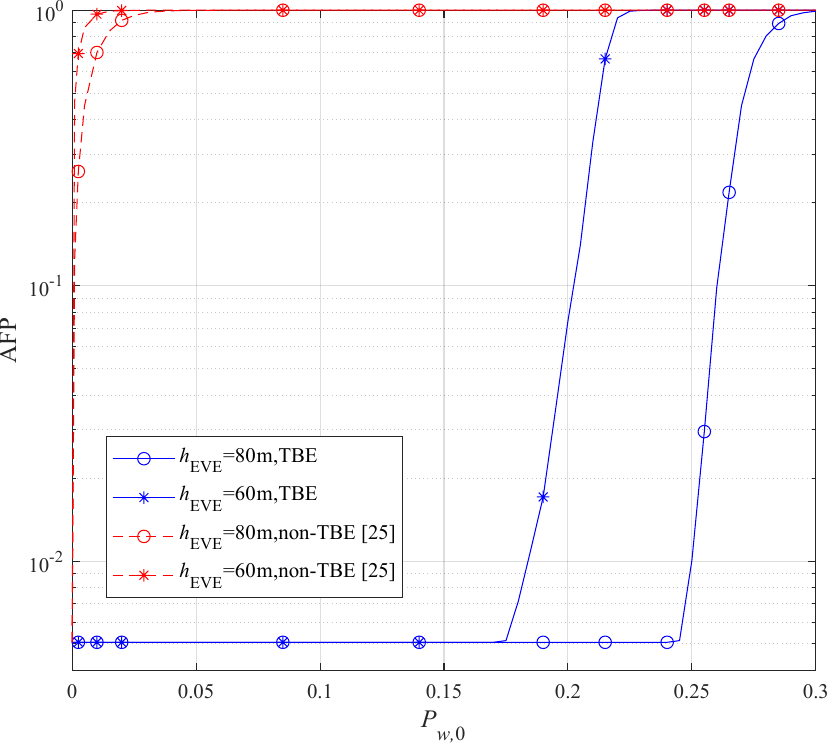}
	\caption{Achievable AFP varying with secrecy constraint $P_{\rm{w},0}$, where $\Delta\theta=1^{\circ}$ and $P_{\rm{d},0}=$ 0.999. 
	In each same scenario, at any prefixed wiretap SER (secrecy constraint), our proposed TBE scheme can achieve lower AFP (better transmission). }\label{afp_com}
\end{figure}

In Fig. \ref{afp_com}, achievable AFP varying with secrecy constraint $P_{\rm{w},0}$ is illustrated, where authentication constraint $P_{\rm{d},0}$ is set to 0.999 and $\Delta\theta$ is fixed as $1^{\circ}$ \cite{UAV}. 
The UAV-EVEs are deployed at an altitude of 60 m or 80 m. 
Obviously, the authentication constraint $P_{\rm{d},0}$ is a little higher than 0.995, which means the global minimum AFP is not achievable. 
The lower bound of achievable AFP is a little higher than the global minimum AFP. 
As the secrecy constraint becomes stricter ($P_{\rm{w},0}$ increasing), all AFP values shift from the unconstrained global optimal value towards 1. 
The more right location of the curve indicates the better transmission performance. 
This rightward shift indicates a lower achievable AFP at a fixed $P_{\rm{w},0}$. 
Additionally, the turning point is determined by the wiretap ability of UAV-EVEs, i.e., the lower $h_{\rm{EVE}}$ is, the turning point occurs at a smaller $P_{\rm{w},0}$. 
The performance of the method from \cite{UAV}, which does not employ TBE, is also provided. 
Our proposed TBE scheme is much better than the non-TBE scheme. 
If $P_{\rm{w},0}$ is set as 0.05 to 0.15, AFP of the TBE scheme is the lower bound while AFP of the non-TBE scheme is 1. 
In addition, whatever $P_{\rm w, 0}$ is, the achievable AFP of TBE scheme is always smaller than or equal to that of prior non-TBE scheme in \cite{UAV}. 
This comparison demonstrates that our proposed TBE scheme can effectively enhance transmission efficiency under secrecy and authentication constraints.

\section{Conclusion}\label{conclusion}
In this paper, we investigated a secrecy rate achieving and authentication enhancement scheme in a chaotic UAV communication environment. 
We proposed a tag-based encoding approach with a novel dual-reference symmetric tag generation mechanism, which reuses the necessary tag for authentication to encode the message, thereby achieving secrecy at a low cost. 
By analyzing the ergodic sum secrecy rate and authentication fail probability, we verified the superiority of our proposed scheme in enhancing security. 
To find suitable power allocations for the tag and artificial noise, two optimization problems were formulated using the above evaluations as objectives, and two corresponding algorithms were provided to solve these problems. 
Our investigation provided a paradigm for pursuing transmission effectiveness and reliability under the security consideration of secrecy and authentication. 
Besides, the balance between the accuracy of tag at users and the fuzziness of tag at eavesdroppers offered a reference for future research on achieving both confidentiality and authentication using limited shared resources. 

\section*{Appendix}
In this section, we prove that $R_{\rm{E}}-R_{\rm U}$ is a difference-of-convex problem and AFP is quasi-convex. 
Specifically, the property of $R_{\rm{sec}}$ is explained by the quasi-concavity of $R_{{\rm E}}$ and $R_{{\rm U}}$. 
Besides, monotonicities are easy to obtain, including $R_{\rm{U}}$ and $R_{\rm{E}}$ increasing with $\phi$, and AFP decreasing with $\phi$. 
Therefore, we prove that $R_{\rm{U}}$ and $R_{\rm{E}}$ are quasi-concave with respect to $\rho$ and AFP are quasi-convex with respect to $\rho$. 

\subsection{The quasi-concavity of $R_{\rm{U}}$}

The sum of concave functions is still concave, so we prove the concavity of $R_u=\log_2\left(1+\text{SINR}_{u,\rm{m}}\right) \cdot P_{u,\rm{d}}$ in (\ref{Rk}). The first-order partial derivation of $R_u$ is 
\begin{equation}
	\begin{array}{l}
	{\partial R_u}/{\partial \rho}=\\
	\log_2\left(1+\text{SINR}_{u,\rm{m}}\right) \cdot \frac{\partial P_{u,\rm{d}}}{\partial \rho}+P_{u,\rm{d}}\cdot \frac{\partial}{\partial \rho} \left(\log_2\left(1+\text{SINR}_{u,\rm{m}}\right)\right).
	\end{array}
\end{equation}
Let ${\partial R_u}/{\partial \rho}=0$, we have 
	\begin{equation}
	\begin{array}{l}
		P_{u,\rm{d}}=-\frac{\partial P_{u,\rm{d}}}{\partial \rho}\cdot
		\frac{\phi\left(1-\rho\right)M + \sigma_{\rm n}^2/{P_{\rm T}\tilde{\beta}}}{\phi M}\ln\left(\frac{\phi M + \sigma_{\rm n}^2/{P_{\rm T}\tilde{\beta}}}{{\phi\left(1-\rho\right)M + \sigma_{\rm n}^2/{P_{\rm T}\tilde{\beta}}}}\right)\\
		=-\frac{\partial P_{u,\rm{d}}}{\partial P_{u,\rm{t}}}\frac{MP_{\rm T}\tilde{\beta}\phi}{2\sigma_{\rm n}^2}
		\frac{\phi\left(1-\rho\right)M + \sigma_{\rm n}^2/{P_{\rm T}\tilde{\beta}}}{\phi M}\ln\left(\frac{\phi M + \sigma_{\rm n}^2/{P_{\rm T}\tilde{\beta}}}{{\phi\left(1-\rho\right)M + \sigma_{\rm n}^2/{P_{\rm T}\tilde{\beta}}}}\right).
	\end{array}\label{Pd_dPd}
\end{equation}
Besides, we have 
\begin{equation}
	\begin{array}{l}
		\frac{\partial P_{u,\rm{d}}}{\partial \rho}=\frac{\partial P_{u,\rm{d}}}{\partial P_{u,\rm{t}}}\frac{\partial P_{u,\rm{t}}}{\partial \rho}=\\
		-A_T^{\eta+1}/A_{\eta}^{\eta}P_{u,\rm{t}}^{\eta-1}\left(1-P_{u,\rm{t}}\right)^{T-\eta-2}\left[P_{u,\rm{t}}\left(1-P_{u,\rm{t}}\right)\right]\frac{\partial P_{u,\rm{t}}}{\partial \rho}. 
	\end{array}\label{dPd}
\end{equation}
Furthermore, we can get 
\begin{equation}
	\begin{array}{l}
		\frac{\partial^2 P_{u,\rm{d}}}{\partial \rho^2}=\frac{\partial^2 P_{u,\rm{d}}}{\partial P_{u,\rm{t}}^2}\left(\frac{\partial P_{u,\rm{t}}}{\partial \rho}\right)^2=-\\
		\frac{A_T^{\eta+1}}{A_{\eta}^{\eta}}P_{u,\rm{t}}^{\eta-1}\left(1-P_{u,\rm{t}}\right)^{T-\eta-2}\left[\eta-\left(T-1\right)P_{u,\rm{t}}\right]\left(\frac{\partial P_{u,\rm{t}}}{\partial \rho}\right)^2.
	\end{array}
\end{equation}
Therefore, 
	\begin{equation}
	\frac{\partial^2 P_{u,\rm{d}}}{\partial \rho^2}=\frac{\eta-\left(T-1\right)P_{u,\rm{t}}}{P_{u,\rm{t}}\left(1-P_{u,\rm{t}}\right)}\frac{\partial P_{u,\rm{t}}}{\partial \rho}\cdot\frac{\partial P_{u,\rm{d}}}{\partial \rho}.\label{d2Pd}
\end{equation}
The second-order partial derivation of $R_u$ can be expressed as 
	\begin{equation}
	\begin{array}{l}
		\frac{\partial^2 R_u}{\partial \rho^2}=\frac{1}{\ln 2}\frac{\phi^2 M^2}{\left(\phi\left(1-\rho\right)M + \sigma_{\rm n}^2/{P_{\rm T}\tilde{\beta}}\right)^2}P_{u,\rm{d}}+\frac{2}{\ln 2}\frac{\phi M\cdot \left(\partial P_{u,\rm{d}}/\partial \rho\right)}{\phi\left(1-\rho\right)M + \sigma_{\rm n}^2/{P_{\rm T}\tilde{\beta}}}\\+\frac{1}{\ln 2}\ln\left(\frac{\phi M + \sigma_{\rm n}^2/{P_{\rm T}\tilde{\beta}}}{{\phi\left(1-\rho\right)M + \sigma_{\rm n}^2/{P_{\rm T}\tilde{\beta}}}}\right) \frac{\partial^2 P_{u,\rm{d}}}{\partial \rho^2}.
	\end{array}\label{d2Rk}
\end{equation}
Substituting (\ref{Pd_dPd}) and (\ref{d2Pd}) into (\ref{d2Rk}), we can get 
\begin{equation}
	\begin{array}{l}
		\frac{\partial^2 R_u}{\partial \rho^2}=	\frac{2}{\ln 2}\frac{\phi M}{\phi\left(1-\rho\right)M + \sigma_{\rm n}^2/{P_{\rm T}\tilde{\beta}}}\cdot \frac{\partial P_{u,\rm{d}}}{\partial \rho}
			\\
			+\frac{1}{\ln 2}\ln\left(\frac{\phi M + \sigma_{\rm n}^2/{P_{\rm T}\tilde{\beta}}}{{\phi\left(1-\rho\right)M + \sigma_{\rm n}^2/{P_{\rm T}\tilde{\beta}}}}\right) \cdot \frac{\eta-\left(T-1\right)P_{u,\rm{t}}}{P_{u,\rm{t}}\left(1-P_{u,\rm{t}}\right)}\frac{\partial P_{u,\rm{d}}}{\partial \rho}\frac{\partial P_{u,\rm{t}}}{\partial \rho}
			\\
			-\frac{1}{\ln 2}\frac{\phi M}{\left(\phi\left(1-\rho\right)M + \sigma_{\rm n}^2/{P_{\rm T}\tilde{\beta}}\right)}\ln\left(\frac{\phi M + \sigma_{\rm n}^2/{P_{\rm T}\tilde{\beta}}}{{\phi\left(1-\rho\right)M + \sigma_{\rm n}^2/{P_{\rm T}\tilde{\beta}}}}\right) \cdot \frac{\partial P_{u,\rm{d}}}{\partial \rho}.
	\end{array}
\end{equation}
For simplification, we have $P_{u,\rm{t}}=Q\left({\sqrt{2MP_{\rm T}\tilde{\beta}}\phi_{\rm s}\rho_{\rm t}}/{\sigma_{\rm n}}\right)$ and $Q\left(x\right)=\frac{1}{2}\exp\left(-\frac{x^2}{2}\right)$ when $x>0$, that is, 
	\begin{equation}
		\begin{array}{l}
			P_{u,\rm{t}}=\frac{1}{2}\exp{\left(MP_{\rm T}\tilde{\beta}\phi\left(\rho-1\right)/\sigma_{\rm n}^2\right)},
			\\
			\frac{\partial  P_{u,\rm{t}}}{\partial \rho}=\frac{MP_{\rm T}\tilde{\beta}\phi}{2\sigma_{\rm n}^2}\exp{\left(MP_{\rm T}\tilde{\beta}\phi\left(\rho-1\right)/\sigma_{\rm n}^2\right)},
		\end{array}\label{Put}
\end{equation}
which means $\frac{\partial P_{u,\rm{t}}}{\partial \rho}=\frac{\phi M}{\sigma_{\rm n}^2/P_{\rm T}\tilde{\beta}}P_{u,\rm{t}}$. 
Observed from (\ref{dPd}) and (\ref{Put}), we have $-\partial P_{u,\rm{d}}/\partial \rho>0$, so 
\begin{equation}
	\begin{array}{*{20}{l}}
		{\frac{{{\partial ^2}{R_u}}}{{\partial {\rho ^2}}}/\left( { - \frac{{\partial {P_{u,\rm{d}}}}}{{\partial \rho }}} \right) \cdot \ln 2 =  - 2\frac{{\phi M}}{{\phi \left( {1 - \rho } \right)M + \sigma_{\rm n}^2/{P_{\rm T}}\tilde \beta }}}\\
		{ - \ln \left( {\frac{{\phi M + \sigma_{\rm n}^2/{P_{\rm T}}\tilde \beta }}{{\phi \left( {1 - \rho } \right)M + \sigma_{\rm n}^2/{P_{\rm T}}\tilde \beta }}} \right)  \frac{{\eta  - \left( {T - 1} \right){P_{u,\rm{t}}}}}{{{P_{u,\rm{t}}}\left( {1 - {P_{u,\rm{t}}}} \right)}}\frac{{\phi M}}{{\sigma_{\rm n}^2/{P_{\rm T}}\tilde \beta }}{P_{u,\rm{t}}}}\\
		{ + \frac{{\phi M}}{{\left( {\phi \left( {1 - \rho } \right)M + \sigma_{\rm n}^2/{P_{\rm T}}\tilde \beta } \right)}}\ln \left( {\frac{{\phi M + \sigma_{\rm n}^2/{P_{\rm T}}\tilde \beta }}{{\phi \left( {1 - \rho } \right)M + \sigma_{\rm n}^2/{P_{\rm T}}\tilde \beta }}} \right)}.
	\end{array}\label{71}
\end{equation}
Above, the first term of (\ref{71}) is negative and the positivity of the other two terms is determined by 
\begin{equation}
	p={\frac{{\phi M}}{{\phi \left( {1 - \rho } \right)M + \sigma_{\rm n}^2/{P_{\rm T}}\tilde \beta }} - \frac{{\eta  - \left( {T - 1} \right){P_{u,\rm{t}}}}}{{1 - {P_{u,\rm{t}}}}}\frac{{\phi M}}{{\sigma_{\rm n}^2/{P_{\rm T}}\tilde \beta }}},
\end{equation}
where $\frac{{\phi M}}{{\phi \left( {1 - \rho } \right)M + \sigma_{\rm n}^2/{P_{\rm T}}\tilde \beta }}<\frac{{\phi M}}{{\sigma_{\rm n}^2/{P_{\rm T}}\tilde \beta }}$. 
Then, observing from (\ref{Pd_dPd}) and (\ref{dPd}), we have 
\begin{equation}
	\begin{array}{l}
		\frac{\partial P_{u,\rm{d}}}{\partial \rho}=-P_{u,\rm{d}} \frac{{2\sigma_{\rm n}^2/{P_{\rm T}}\tilde \beta }}{{\phi M}}
		\exp^{-1}{\left(MP_{\rm T}\tilde{\beta}\phi\left(\rho-1\right)/\sigma_{\rm n}^2\right)}
		\\ \cdot
		\frac{{\phi M}}{{\phi \left( {1 - \rho } \right)M + \sigma_{\rm n}^2/{P_{\rm T}}\tilde \beta }}{\ln ^{ - 1}}\left( {\frac{{\phi M + \sigma_{\rm n}^2/{P_{\rm T}}\tilde \beta }}{{\phi \left( {1 - \rho } \right)M + \sigma_{\rm n}^2/{P_{\rm T}}\tilde \beta }}} \right).
	\end{array}
\end{equation}
Assuming that the tag is covert and is comparable to the noise, we have  $\exp^{-1}{\left(MP_{\rm T}\tilde{\beta}\phi\left(\rho-1\right)/\sigma_{\rm n}^2\right)}$ is close to $\exp 1$, and $\frac{{2\sigma_{\rm n}^2/{P_{\rm T}}\tilde \beta }}{{\phi M}}\frac{{\phi M}}{{\phi \left( {1 - \rho } \right)M + \sigma_{\rm n}^2/{P_{\rm T}}\tilde \beta }}$ is close to $1$, but ${\ln ^{ - 1}}\left( {\frac{{\phi M + \sigma_{\rm n}^2/{P_{\rm T}}\tilde \beta }}{{\phi \left( {1 - \rho } \right)M + \sigma_{\rm n}^2/{P_{\rm T}}\tilde \beta }}} \right)$ is much smaller than other terms. 
Besides, $P_{u,\rm{d}}$ is in $(0,1)$, so we have $|\frac{\partial P_{u,\rm{d}}}{\partial \rho}|_{\frac{\partial R_u}{\partial \rho}=0}|$ is not large. 
On the other hand, when $\frac{\partial^2 P_{u,\rm{d}}}{\partial \rho^2}=0$ ($P_{u,\rm{t}}=\frac{\eta}{T-1}$), the varying of $\frac{\partial P_{u,\rm{d}}}{\partial \rho}$ is the fastest and $|\frac{\partial P_{u,\rm{d}}}{\partial \rho}|$ is the largest (much larger than $1$). 
In addition, we have $P_{u,\rm{t}}|_{\frac{{\eta  - \left( {T - 1} \right){P_{u,\rm{t}}}}}{{1 - {P_{u,\rm{t}}}}}=1}=\frac{\eta-1}{T-2}$, which is close to $P_{u,\rm{t}}|_{\eta-\left(T-1\right)P_{u,\rm{t}}=0}=\frac{\eta}{T-1}$. 
Above, when ${\partial R_k}/{\partial \rho}=0$, we have $P_{u,\rm{t}}<<\frac{\eta}{T-1}$, that is, $\frac{{\eta  - \left( {T - 1} \right){P_{u,\rm{t}}}}}{{1 - {P_{u,\rm{t}}}}}$ is larger than $1$. 
Therefore, $p$ is negative when ${\partial R_u}/{\partial \rho}=0$. 
Thus, 
\begin{equation}
	\frac{\partial^2 R_u}{\partial \rho^2}|_{{\partial R_u}/{\partial \rho}=0}<0,
\end{equation}
which means $R_u$ is quasi-concave and $R_{\rm U}$ is quasi-concave.

\subsection{The quasi-concavity of $R_{\rm{E}}$}
Similarly, we focus on proving the concavity of $R_e^{'}=\log_2\left(1+\text{SINR}_{e,\rm{m}}^{'}\right) \cdot I_e$, where $I_e=1+\log_2\left(1-P_{e,\rm{t}}^{'}\right)$. 
The first-order derivation is 
\begin{equation}
	\begin{array}{l}
			\frac{\partial R_e^{'}}{\partial \rho}
			=
			\frac{\phi\beta_u^{-1}\tilde{\beta}f\left(\kappa\right)/\ln2}{\phi\left(1-\rho\right)\beta_u^{-1}\tilde{\beta}f\left(\kappa\right)+\left(1-\phi\right)g\left(\kappa\right)+{\sigma_{\rm n}^2}/{P_{\rm T}\alpha_e}}I_e
			\\
			-\frac{1}{\ln2}\ln\left(\frac{\phi\beta_u^{-1}\tilde{\beta}f\left(\kappa\right)+\left(1-\phi\right)g\left(\kappa\right)+{\sigma_{\rm n}^2}/{P_{\rm T}\alpha_e}}{\phi\left(1-\rho\right)\beta_u^{-1}\tilde{\beta}f\left(\kappa\right)+\left(1-\phi\right)g\left(\kappa\right)+{\sigma_{\rm n}^2}/{P_{\rm T}\alpha_e}}\right)
		\frac{\partial I_e}{\partial \rho}.
	\end{array}
\end{equation}
Let ${\partial R_e^{'}}/{\partial \rho}=0$, we have 
\begin{equation}
	\begin{array}{l}
		I_e=\frac{\phi\left(1-\rho\right)\beta_u^{-1}\tilde{\beta}f\left(\kappa\right)+\left(1-\phi\right)g\left(\kappa\right)+{\sigma_{\rm n}^2}/{P_{\rm T}\alpha_e}}{\phi\beta_u^{-1}\tilde{\beta}f\left(\kappa\right)}\cdot
		\\
		\ln\left(\frac{\phi\beta_u^{-1}\tilde{\beta}f\left(\kappa\right)+\left(1-\phi\right)g\left(\kappa\right)+{\sigma_{\rm n}^2}/{P_{\rm T}\alpha_e}}{\phi\left(1-\rho\right)\beta_u^{-1}\tilde{\beta}f\left(\kappa\right)+\left(1-\phi\right)g\left(\kappa\right)+{\sigma_{\rm n}^2}/{P_{\rm T}\alpha_e}}\right)
		\frac{\partial I_e}{\partial \rho}.
	\end{array}
\end{equation}
Besides, we have 
			$\frac{\partial I_e}{\partial \rho}=\frac{-1}{1-P_{e,\rm{t}}^{'}}\cdot \frac{\partial P_{e,\rm{t}}^{'}}{\partial \rho}$ and 
\begin{equation}
	\begin{array}{l}
			\frac{\partial^2 I_e}{\partial \rho^2}=\frac{\partial^2 I_e}{\left(\partial P_{e,\rm{t}}^{'}\right)^2}\left(\frac{\partial P_{e,\rm{t}}^{'}}{\partial \rho}\right)^2=\frac{-1}{\left(1-P_{e,\rm{t}}\right)^2}\left(\frac{\partial P_{e,\rm{t}}^{'}}{\partial \rho}\right)^2
   <0.
	\end{array}
\end{equation}
The second-order derivation is 
\begin{equation}
	\begin{array}{l}
		\frac{\partial^2 R_e^{'}}{\partial \rho^2}=\frac{1}{\ln 2}I_e
		\frac{-\phi\beta_u^{-1}\tilde{\beta}f\left(\kappa\right)\cdot\left(-\phi\beta_u^{-1}\tilde{\beta}f\left(\kappa\right)\right) }{\left(\phi\left(1-\rho\right)\beta_u^{-1}\tilde{\beta}f\left(\kappa\right)+\left(1-\phi\right)g\left(\kappa\right)+{\sigma_{\rm n}^2}/{P_{\rm T}\alpha_e}\right)^2}
		\\
		+\frac{2}{\ln 2}\frac{\partial I_e}{\partial \rho}
		\frac{\phi\beta_u^{-1}\tilde{\beta}f\left(\kappa\right)}{\left(\phi\left(1-\rho\right)\beta_u^{-1}\tilde{\beta}f\left(\kappa\right)+\left(1-\phi\right)g\left(\kappa\right)+{\sigma_{\rm n}^2}/{P_{\rm T}\alpha_e}\right)}
		\\
		+\frac{1}{\ln 2}\ln\left(\frac{\phi\beta_u^{-1}\tilde{\beta}f\left(\kappa\right)+\left(1-\phi\right)g\left(\kappa\right)+{\sigma_{\rm n}^2}/{P_{\rm T}\alpha_e}}{\phi\left(1-\rho\right)\beta_u^{-1}\tilde{\beta}f\left(\kappa\right)+\left(1-\phi\right)g\left(\kappa\right)+{\sigma_{\rm n}^2}/{P_{\rm T}\alpha_e}}\right)
		\frac{\partial^2 I_e}{\partial \rho^2}.
	\end{array}
\end{equation}
For simplicity, we have $
		P_{e,\rm{t}}^{'}=\frac{1}{2}\exp\left(\frac{\left(\rho-1\right)\phi_{\rm s}^2\beta_u^{-1}\tilde{\beta}f\left(\kappa\right)}{\sigma_{\rm n}^2/P_{\rm T}\alpha_e+\phi_{\rm n}^2g\left(\kappa\right)}\right)>0$ and 
\begin{equation}
	\begin{array}{l}
		\frac{dP_{e,\rm{t}}^{'}}{d\rho}=\frac{1}{2}\frac{\phi\beta_u^{-1}\tilde{\beta}f\left(\kappa\right)}{2\sigma_{\rm n}^2/P_{\rm T}\alpha_e+2\left(1-\phi\right)g\left(\kappa\right)}\exp\left(\frac{\left(\rho-1\right)\phi_{\rm s}^2\beta_u^{-1}\tilde{\beta}f\left(\kappa\right)}{\sigma_{\rm n}^2/P_{\rm T}\alpha_e+\phi_{\rm n}^2g\left(\kappa\right)}\right)>0. 
	\end{array}
\end{equation}
The second-order derivation can be written as 
\begin{equation}
	\begin{array}{*{20}{l}}
		\begin{array}{l}
			\frac{{{\partial ^2}R_e^{'}}}{{\partial {\rho ^2}}} = \frac{1}{{\ln 2}}\frac{{\phi \beta _u^{ - 1}\tilde \beta f\left( \kappa  \right)}}{{\phi \left( {1 - \rho } \right)\beta _u^{ - 1}\tilde \beta f\left( \kappa  \right) + \left( {1 - \phi } \right)g\left( \kappa  \right) + \sigma_{\rm n}^2/{P_{\rm T}}{\alpha _e}}} \cdot \\
			\ln \left( {\frac{{\phi \beta _u^{ - 1}\tilde \beta f\left( \kappa  \right) + \left( {1 - \phi } \right)g\left( \kappa  \right) + \sigma_{\rm n}^2/{P_{\rm T}}{\alpha _e}}}{{\phi \left( {1 - \rho } \right)\beta _u^{ - 1}\tilde \beta f\left( \kappa  \right) + \left( {1 - \phi } \right)g\left( \kappa  \right) + \sigma_{\rm n}^2/{P_{\rm T}}{\alpha _e}}}} \right)\frac{{\partial {I_e}}}{{\partial \rho }}
		\end{array}\\
		{ + \frac{2}{{\ln 2}}\frac{{\partial {I_e}}}{{\partial \rho }}\frac{{\phi \beta _u^{ - 1}\tilde \beta f\left( \kappa  \right)}}{{\left( {\phi \left( {1 - \rho } \right)\beta _u^{ - 1}\tilde \beta f\left( \kappa  \right) + \left( {1 - \phi } \right)g\left( \kappa  \right) + \sigma_{\rm n}^2/{P_{\rm T}}{\alpha _e}} \right)}}}\\
		{ - \frac{1}{{\ln 2}}\ln \left( {\frac{{\phi \beta _u^{ - 1}\tilde \beta f\left( \kappa  \right) + \left( {1 - \phi } \right)g\left( \kappa  \right) + \sigma_{\rm n}^2/{P_{\rm T}}{\alpha _e}}}{{\phi \left( {1 - \rho } \right)\beta _u^{ - 1}\tilde \beta f\left( \kappa  \right) + \left( {1 - \phi } \right)g\left( \kappa  \right) + \sigma_{\rm n}^2/{P_{\rm T}}{\alpha _e}}}} \right)\frac{{{\partial ^2}{I_e}}}{{{\partial ^2}\rho }}}
	\end{array}
\end{equation}
where $\partial I_e/\partial \rho<0$.
Thus, $\frac{{{\partial ^2}R_e^{'}}}{{\partial {\rho ^2}}}|_{\frac{\partial R_{e}^{'}}{\partial \rho}=0}<0$, that is, $R_e^{'}$ is quasi-concave and $R_{\rm E}$ is quasi-concave. 

\subsection{The quasi-convexity of \rm{AFP}}
\label{AFP_convex}
We have 
\begin{equation}
	\text{AFP}=1-\left(1-P_{u,\rm{m}}\right)^T\cdot P_{u,\rm{d}}. 
\end{equation}
The first-order derivation can be written as
\begin{equation}
	\begin{array}{l}
		\frac{\partial \text{AFP}}{\partial \rho}=-T\cdot \left(1-P_{u,\rm{m}}\right)^{T-1}\cdot \left(-\frac{\partial P_{u,\rm{m}}}{\partial \rho}\right)\cdot P_{u,\rm{d}}
		\\
		-\left(1-P_{u,\rm{m}}\right)^T\cdot\frac{\partial P_{u,\rm{d}}}{\partial P_{u,\rm{t}}}\cdot\frac{\partial P_{u,\rm{t}}}{\partial \rho}.
	\end{array}
\end{equation}
Let $\partial \text{AFP}/\partial \rho=0$, we have 
	\begin{equation}
		P_{u,\rm{d}}=\frac{1}{T}\left(1-P_{u,\rm{m}}\right)\cdot\frac{\partial P_{u,\rm{d}}}{\partial P_{u,\rm{t}}}\cdot\frac{\partial P_{u,\rm{t}}}{\partial \rho}/ \left(\frac{\partial P_{u,\rm{m}}}{\partial \rho}\right).\label{Pd_AFP}
\end{equation}
The second-order derivation is 
\begin{equation}
	\begin{array}{l}
		\frac{\partial^2 \text{AFP}}{\partial \rho^2}=
		-T\left(T-1\right)\left(1-P_{u,\rm{m}}\right)^{T-2}\left(-\frac{\partial P_{u,\rm{m}}}{\partial \rho}\right)^2\cdot P_{u,\rm{d}}
		\\
		-T\left(1-P_{u,\rm{m}}\right)^{T-1}\left(-\frac{\partial^2 P_{u,\rm{m}}}{\partial \rho^2}\right)\cdot P_{u,\rm{d}}
		\\
		-T\cdot \left(1-P_{u,\rm{m}}\right)^{T-1}\cdot \left(-\frac{\partial P_{u,\rm{m}}}{\partial \rho}\right)\cdot \frac{\partial P_{u,\rm{d}}}{\partial P_{u,\rm{t}}}\frac{\partial P_{u,\rm{t}}}{\partial \rho}
		\\
		-T\left(1-P_{u,\rm{m}}\right)^{T-1}\cdot\left(-\frac{\partial P_{u,\rm{m}}}{\partial \rho}\right)\cdot\frac{\partial P_{u,\rm{d}}}{\partial P_{u,\rm{t}}}\cdot\frac{\partial P_{u,\rm{t}}}{\partial \rho}
		\\
		-\left(1-P_{u,\rm{m}}\right)^T\cdot\frac{\partial^2 P_{u,\rm{d}}}{\partial P_{u,\rm{t}}^2}\cdot\frac{\partial P_{u,\rm{t}}}{\partial \rho}\cdot\frac{\partial P_{u,\rm{t}}}{\partial \rho}
		\\
		-\left(1-P_{u,\rm{m}}\right)^T\cdot\frac{\partial P_{u,\rm{d}}}{\partial P_{u,\rm{t}}}\cdot\frac{\partial^2 P_{u,\rm{t}}}{\partial \rho^2}.
	\end{array}\label{erjie}
\end{equation}
Substituting (\ref{Pd_AFP}) into (\ref{erjie}), we have 
\begin{equation}
	\begin{array}{l}
		\frac{\partial^2 \text{AFP}}{\partial \rho^2}=\\
		-\left(T-1\right)\left(1-P_{u,\rm{m}}\right)^{T-1}\left(\frac{\partial P_{u,\rm{m}}}{\partial \rho}\right)^2\frac{\partial P_{u,\rm{d}}}{\partial P_{u,\rm{t}}}\cdot\frac{\partial P_{u,\rm{t}}}{\partial \rho}/ \frac{\partial P_{u,\rm{m}}}{\partial \rho}
		\\
		-\left(1-P_{u,\rm{m}}\right)^{T}\left(-\frac{\partial^2 P_{u,\rm{m}}}{\partial \rho^2}\right)\frac{\partial P_{u,\rm{d}}}{\partial P_{u,\rm{t}}}\cdot\frac{\partial P_{u,\rm{t}}}{\partial \rho}/ \frac{\partial P_{u,\rm{m}}}{\partial \rho}
		\\
		-2T\cdot \left(1-P_{u,\rm{m}}\right)^{T-1}\cdot \left(-\frac{\partial P_{u,\rm{m}}}{\partial \rho}\right)\cdot \frac{\partial P_{u,\rm{d}}}{\partial P_{u,\rm{t}}}\frac{\partial P_{u,\rm{t}}}{\partial \rho}
		\\
		-\left(1-P_{u,\rm{m}}\right)^T\cdot\frac{\partial^2 P_{u,\rm{d}}}{\partial P_{u,\rm{t}}^2}\cdot\frac{\partial P_{u,\rm{t}}}{\partial \rho}\cdot\frac{\partial P_{u,\rm{t}}}{\partial \rho}
		\\
		-\left(1-P_{u,\rm{m}}\right)^T\cdot\frac{\partial P_{u,\rm{d}}}{\partial P_{u,\rm{t}}}\cdot\frac{\partial^2 P_{u,\rm{t}}}{\partial \rho^2}.
	\end{array}
\end{equation}
Furthermore, we have 
\begin{equation}
	\begin{array}{l}
		\frac{\partial^2 \text{AFP}}{\partial \rho^2}/\left(-\frac{\partial P_{u,\rm{d}}}{\partial P_{u,\rm{t}}}\right)/\left(1-P_{u,\rm{m}}\right)^{T-1}=\\
		-\left(T+1\right)\frac{\partial P_{u,\rm{m}}}{\partial \rho}\frac{\partial P_{u,\rm{t}}}{\partial \rho}
		-\frac{\partial^2 P_{u,\rm{m}}}{\partial \rho^2}\left(1-P_{u,\rm{m}}\right)\frac{\partial P_{u,\rm{t}}}{\partial \rho}/ \frac{\partial P_{u,\rm{m}}}{\partial \rho}
		\\
		+\left(1-P_{u,\rm{m}}\right)\frac{\eta-\left(T-1\right)P_{u,\rm{t}}}{P_{u,\rm{t}}\left(1-P_{u,\rm{t}}\right)}\frac{P_{u,\rm{t}}}{\partial \rho}\frac{\partial P_{u,\rm{t}}}{\partial \rho}
		+\left(1-P_{u,\rm{m}}\right)\frac{\partial^2 P_{u,\rm{t}}}{\partial \rho^2}.
	\end{array}\label{afp2}
\end{equation}
Besides, 
\begin{equation}
	\begin{array}{l}
 P_{u,\rm{t}}=\frac{1}{2}\exp{\left(MP_{\rm T}\tilde{\beta}\phi\left(\rho-1\right)/\sigma_{\rm n}^2\right)},
	\end{array}\label{pt0}
\end{equation}
\begin{equation}
	\begin{array}{l}
			\frac{\partial  P_{u,\rm{t}}}{\partial \rho}=\frac{MP_{\rm T}\tilde{\beta}\phi}{2\sigma_{\rm n}^2}\exp{\left(MP_{\rm T}\tilde{\beta}\phi\left(\rho-1\right)/\sigma_{\rm n}^2\right)},
	\end{array}\label{pt1}
\end{equation}
\begin{equation}
	\begin{array}{l}
		\frac{\partial^2P_{u,\rm{t}}}{\partial\rho^2}=\frac{M^2P_{\rm T}^2{\tilde{\beta}}^2\phi^2}{2\sigma_n^4}\exp{\left(MP_{\rm T}\tilde{\beta}\phi\left(\rho-1\right)/\sigma_{\rm n}^2\right)},
	\end{array}\label{pt2}
\end{equation}
\begin{equation}
	\begin{array}{l}
		P_{u,\rm{m}}=\exp\left(-\frac{{MP_{\rm T}\tilde{\beta}}\phi\rho}{2\sigma_{\rm n}^2}\right)-\frac{1}{4}\exp^2\left(-\frac{{MP_{\rm T}\tilde{\beta}}\phi\rho}{2\sigma_{\rm n}^2}\right),
	\end{array}\label{pm0}
\end{equation}
\begin{equation}
	\begin{array}{l}
		\frac{\partial P_{u,\rm{m}}}{\partial \rho}
		=-\frac{MP_{\rm T}\tilde{\beta}\phi}{4\sigma_{\rm n}^2}\exp\left(-MP_{\rm T}\tilde{\beta}\phi\rho/2\sigma_{\rm n}^2\right),
	\end{array}\label{pm1}
\end{equation}
\begin{equation}
	\begin{array}{l}
		\frac{\partial^2 P_{u,\rm{m}}}{\partial \rho^2}=\frac{M^2P_{\rm T}^2\tilde{\beta}^2\phi^2}{8\sigma_n^4}\exp\left(-MP_{\rm T}\tilde{\beta}\phi\rho/2\sigma_{\rm n}^2\right).
	\end{array}\label{pm2}
\end{equation}

Let  $ET=\exp{\left(MP_{\rm T}\tilde{\beta}\phi\left(\rho-1\right)/\sigma_{\rm n}^2\right)}$ and $EM=$ $\exp\left(-MP_{\rm T}\tilde{\beta}\phi\rho/2\sigma_{\rm n}^2\right)$.
Substituting (\ref{pt0}), (\ref{pt1}), (\ref{pt2}), (\ref{pm0}), (\ref{pm1}), and (\ref{pm2}) into (\ref{afp2}), we have
\begin{equation}
	\begin{array}{l}
		\frac{\partial^2 \text{AFP}}{\partial \rho^2}/\left(-\frac{\partial P_{u,\rm{d}}}{\partial P_{u,\rm{t}}}\right)/\left(1-P_{u,\rm{m}}\right)^{T-1}=\\
		-\left(T+1\right)\left(-\frac{MP_{\rm T}\tilde{\beta}\phi}{4\sigma_{\rm n}^2}EM\right)\frac{MP_{\rm T}\tilde{\beta}\phi}{2\sigma_{\rm n}^2}ET
		\\
		-\frac{M^2P_{\rm T}^2\tilde{\beta}^2\phi^2}{8\sigma_n^4}EM\left(1-P_{u,\rm{m}}\right)\frac{MP_{\rm T}\tilde{\beta}\phi}{2\sigma_{\rm n}^2}ET/ \left(-\frac{MP_{\rm T}\tilde{\beta}\phi}{4\sigma_{\rm n}^2}EM\right)
		\\
		+\left(1-P_{u,\rm{m}}\right)\frac{\eta-\left(T-1\right)P_{u,\rm{t}}}{P_{u,\rm{t}}\left(1-P_{u,\rm{t}}\right)}\frac{MP_{\rm T}\tilde{\beta}\phi}{2\sigma_{\rm n}^2}ET\frac{MP_{\rm T}\tilde{\beta}\phi}{2\sigma_{\rm n}^2}ET
		\\
		+\left(1-P_{u,\rm{m}}\right)\frac{M^2P_{\rm T}^2{\tilde{\beta}}^2\phi^2}{2\sigma_n^4}ET.
	\end{array}
\end{equation}
After calculation, we have 
\begin{equation}
	\begin{array}{l}
		\frac{\partial^2 \text{AFP}}{\partial \rho^2}/\left(-\frac{\partial P_{u,\rm{d}}}{\partial P_{u,\rm{t}}}\right)/\left(1-P_{u,\rm{m}}\right)^{T-1}/ET/\frac{M^2P_T^2\tilde{\beta}^2\phi^2}{8\sigma_n^4}\\
		=\left(T+1\right)EM
		+2\left(1-P_{u,\rm{m}}\right)+4\left(1-P_{u,\rm{m}}\right)
		\\
		+\left(1-P_{u,\rm{m}}\right)\frac{\eta-\left(T-1\right)P_{u,\rm{t}}}{P_{u,\rm{t}}\left(1-P_{u,\rm{t}}\right)}2ET.
	\end{array}
\end{equation}

We can confirm that $P_{u,\rm{t}}<\eta/\left(T-1\right)$ when $\partial \text{AFP}/\partial \rho=0$. 
When $P_{u,\rm{t}}=\eta/\left(T-1\right)$, the decrease speed of $P_{u,\rm{d}}$ is the fastest. 
When $\rho$ left approaches this point, the decrease of $P_{u,\rm{d}}$ is faster than the increase of $\left(1-P_{u,\rm{t}}\right)^T$, that is, $\left(1-P_{u,\rm{t}}\right)^TP_{u,\rm{d}}$ decreases and AFP increases. 
It means $\rho|_{\partial \text{AFP}/\partial \rho=0}<\rho|_{\partial^2 P_{u,\rm{d}}/\partial \rho^2=0}$. 
Thus, $\partial \text{AFP}/\partial \rho=0$ is achieved when $P_{u,\rm{t}}<\eta/\left(T-1\right)$, so $\eta/P_{u,\rm{t}}+1-T>0$, that is, we have 
\begin{equation}
	\frac{\partial^2 \text{AFP}}{\partial \rho^2}|_{\frac{\partial \text{AFP}}{\partial \rho}=0}>0.
\end{equation}

Above, AFP is quasi-convex.


\begin{thebibliography}{00}
\bibitem{Tutorial1}
L. Gupta, R. Jain, and G. Vaszkun, ``Survey of Important Issues in UAV Communication Networks," \emph{IEEE Commun. Surv. Tutor.}, vol. 18, no. 2, pp. 1123-1152, Second quarter 2016. 
\bibitem{Tutorial2}
M. Mozaffari, W. Saad, M. Bennis, Y. Nam, and M. Debbah, ``A Tutorial on UAVs for Wireless Networks: Applications, Challenges, and Open Problems," \emph{IEEE Commun. Surv. Tutor.}, vol. 21, no. 3, pp. 2334-2360, third quarter 2019. 
\bibitem{Tutorial3}
A. Fotouhi, Y. Chen, N. Zhao, M. Alouini, and P. Dobbins, ``Survey on UAV Cellular Communications: Practical Aspects, Standardization Advancements, Regulation, and Security Challenges," \emph{IEEE Commun. Surv. Tutor.}, vol. 21, no. 4, pp. 3417-3442, Fourth quarter 2019. 
 \bibitem{secrecy}
 G. Pan, H. Lei, J. An, S. Zhang, and M. Alouini, "On the Secrecy of UAV Systems With Linear Trajectory," \emph{IEEE Trans. Wireless Commun.}, vol. 19, no. 10, pp. 6277-6288, Oct. 2020. 
 \bibitem{authentication}
 C. Jiang, Y. Fang, P. Zhao, and J. Panneerselvam, ``Intelligent UAV Identity Authentication and Safety Supervision Based on Behavior Modeling and Prediction," \emph{IEEE Trans Industr Inform}, vol. 16, no. 10, pp. 6652-6662, Oct. 2020. 
\bibitem{PLS1}
J. M. Hamamreh, H. M. Furqan, and H. Arslan, ``Classifications and Applications of Physical Layer Security Techniques for Confidentiality: A Comprehensive Survey," \emph{IEEE Commun. Surv. Tutor.}, vol. 21, no. 2, pp. 1773-1828, Second quarter 2019. 
 \bibitem{PLS2}
 A. K. Yerrapragada, T. Eisman, and B. Kelley, ``Physical Layer Security for Beyond 5G: Ultra Secure Low Latency Communications," \emph{IEEE Open J. Commun. Soc.}, vol. 2, pp. 2232-2242, 2021. 
 \bibitem{low_latency}
 G. K. Pandey, D. S. Gurjar, H. H. Nguyen and S. Yadav, ``Security Threats and Mitigation Techniques in UAV Communications: A Comprehensive Survey," \emph{IEEE Access}, vol. 10, pp. 112858-112897, 2022. 
 \bibitem{supplement}
 H. Wang, H. Fang, and X. Wang, ``Safeguarding Cluster Heads in UAV Swarm Using Edge Intelligence: Linear Discriminant Analysis-Based Cross-Layer Authentication," \emph{IEEE Open J. Commun. Soc.}, vol. 2, pp. 1298-1309, 2021. 
 	\bibitem{UAV-friendly jamming STAR-RIS}
	A. M. Benaya, M. H. Ismail, A. S. Ibrahim, and A. A. Salem, ``Physical Layer Security Enhancement via Intelligent Omni-Surfaces and UAV-Friendly Jamming," \emph{IEEE ACCESS}, vol. 11, pp. 2531-2544, Jan. 2023. 
	\bibitem{UAV-friendly jamming UAV NOMA}
	D. Diao, B. Wang, K. Cao, R. Dong, and T. Cheng, ``Enhancing Reliability and Security of UAV-Enabled NOMA Communications with Power Allocation and Aerial Jamming," \emph{IEEE Trans. Veh. Technol.}, vol. 71, no. 8, pp. 8662-8674, Aug. 2022. 
 	\bibitem{UAV_relay_beamforming}
	R. Dong, B. Wang, and K. Cao, ``Security Enhancement of UAV Swarm Enabled Relaying Systems with Joint Beamforming and Resource Allocation," \emph{China Commun.}, vol. 18, no. 9, pp. 71-87, Sept. 2021. 
	\bibitem{encryption}
	A. Maksud and Y. Hua, ``Physical Layer Encryption for UAV-to-Ground Communications," \emph{IEEE Int. Conf. Commun. (ICC Workshops)}, Seoul, Korea, Republic of, 2022, pp. 1077-1082. 
 \bibitem{key_generation}
 S. Jangsher, A. Al-Dweik, Y. Iraqi, A. Pandey, and J. Giacalone, ``Group Secret Key Generation Using Physical Layer Security for UAV Swarm Communications," \emph{IEEE Trans. Aerosp. Electron. Syst.}, vol. 59, no. 6, pp. 8550-8564, Dec. 2023.
 	\bibitem{UAV_transmitter_AN_beamforming}
 	Y. Chen, G. Liu, Z. Zhang, L. He, and S. He, ``Improving Physical Layer Security for multi-UAV Systems Against Hybrid Wireless Attacks," \emph{IEEE Trans. Veh. Technol.}, vol. 73, no. 5, pp. 7034-7048, May 2024. 


	\bibitem{authentication_Xie}
	N. Xie, Z. Li, and H. Tan, ``A Survey of Physical-Layer Authentication in Wireless Communications," \emph{IEEE Commun. Surv. Tutor.}, vol. 23, no. 1, pp. 282-310, Dec. 2021.
	\bibitem{CSI}
	J. Liu and X. Wang, ``Physical Layer Authentication Enhancement Using Two-Dimensional Channel Quantization'', \emph{IEEE Trans. Wireless Commun.}, vol. 15, no. 6, pp. 4171-4182, Jun. 2016. 
	\bibitem{CSI2}
	X. Lu, J. Lei, Y. Shi, and W. Li, ``Physical-Layer Authentication Based on Channel Phase Responses for Multi-Carriers Transmission," \emph{IEEE Trans. Inf. Forensics Security}, vol. 18, pp. 1734-1748, March 2023. 
	\bibitem{PLA_CSI}
	Y. Zhou, Z. Ma, H. Liu, P. L. Yeoh, Y. Li, B. Vucetic, and P. Fan, ``A UAV-Aided Physical Layer Authentication Based on Channel Characteristics and Geographical Locations," \emph{IEEE Trans. Veh. Technol.}, vol. 73, no. 1, pp. 1053-1064, Jan. 2024. 
	\bibitem{CSI_CFO_PN}
	Y. Teng, P. Zhang, Y. Liu, J. Dong, and F. Xiao, ``Exploiting Carrier Frequency Offset and Phase Noise for Physical Layer Authentication in UAV-aided Communication Systems," \emph{IEEE Trans. Commun.}, vol. 72, no. 8, pp. 4708-4724, Aug. 2024. 
	\bibitem{UAV_conference}
	S. J. Maeng, Y. Yapici,˙I. Güvenç, H. Dai, and A. Bhuyan, "Precoder design for mmWave UAV communications with physical layer security," \emph{Proc. IEEE Sig. Proc. Adv. Wireless Commun.}, Atlanta, GA, 2020, pp. 1–5. 
	\bibitem{Paul_part1}
	J. B. Perazzone, E. Graves, P. L. Yu, and R. S. Blum, ``Secret Key-Enabled Authenticated-Capacity Region, Part I: Average Authentication,'' \emph{"IEEE Trans. Inform. Theory}, vol. 68, no. 10, pp. 6802-6825, Oct. 2022. 
	\bibitem{Paul_part2}
	E. Graves, J. B. Perazzone, Paul L. Yu, and R. S. Blum, ``Secret key-enabled authenticated-capacity region, Part—II: Typical-authentication,'' \emph{"IEEE Trans. Inform. Theory}, vol. 68, no. 11, pp. 6981-7004, Nov. 2022. 
	 \bibitem{spatial correlation property}
	Y. Teng, P. Zhang, X. Chen, X. Jiang, and F. Xiao, "PHY-Layer Authentication Exploiting Channel Sparsity in MmWave MIMO UAV-Ground Systems," \emph{IEEE Trans. Inf. Forensics Security}, vol. 19, pp. 4642-4657, 2024. 
 \bibitem{UAV}
 S. J. Maeng, A. Bhuyan, and H. Dai, ``Precoder Design for Physical-Layer Security and Authentication in Massive MIMO UAV Communications," \emph{IEEE Trans. Veh. Technol.}, vol. 71, no. 3, pp. 2949-2964, Mar. 2022. 
 \bibitem{blockchain1}
 W. Wang, Z. Han, T. R. Gadekallu, S. Raza, J. Tanveer, and C. Su, ``Lightweight Blockchain-Enhanced Mutual Authentication Protocol for UAVs," \emph{IEEE Internet Things J.}, vol. 11, no. 6, pp. 9547-9557, 15 Mar., 2024. 
 \bibitem{blockchain2}
 Y. Tan, J. Wang, J. Liu, and N. Kato, ``Blockchain-Assisted Distributed and Lightweight Authentication Service for Industrial Unmanned Aerial Vehicles," \emph{IEEE Internet Things J.}, vol. 9, no. 18, pp. 16928-16940, 15 Sept., 2022. 
 \bibitem{blockchanin3}
 R. Karmakar, G. Kaddoum and O. Akhrif, ``A Blockchain-Based Distributed and Intelligent Clustering-Enabled Authentication Protocol for UAV Swarms," \emph{IEEE Transactions on Mobile Computing}, vol. 23, no. 5, pp. 6178-6195, May 2024. 
 \bibitem{UAV channel}
 P. S. Bithas, V. Nikolaidis, A. G. Kanatas, and G. K. Karagiannidis, ``UAV-to-Ground Communications: Channel Modeling and UAV Selection," \emph{IEEE Trans. Commun.}, vol. 68, no. 8, pp. 5135-5144, Aug. 2020. 
 \bibitem{risk}
 H. Shakhatreh, A. Sawalmeh, K. F. Hayajneh, S. Abdel-Razeq, W. Malkawi, and A. Al-Fuqaha, ``A Systematic Review of Interference Mitigation Techniques in Current and Future UAV-Assisted Wireless Networks,” \emph{IEEE Open J. Commun. Soc.}, vol. 5, pp. 2815-2846, Apr. 2024.
 \bibitem{encryption2}
 T. Assaf, A. Arafat, Y. Iraqi, S. Jangsher, A. Pandey, J. Giacalone, E. Abulibdeh, H. Saleh, and B. Mohammad, ``High-Rate Secret Key Generation Using Physical Layer Security and Physical Unclonable Functions," \emph{IEEE Open J. Commun. Soc.}, vol. 4, pp. 209-225, 2023. 
\bibitem{tag_RIS}
P. Zhang, Y. Teng, Y. Shen, X. Jiang, and F. Xiao, ``Tag-Based PHY-Layer Authentication for RIS-Assisted Communication Systems," \emph{IEEE Trans. Dependable Secure Comput.}, vol. 20, no. 6, pp. 4778-4792, Nov.-Dec. 2023. 
\bibitem{tag_Ning_Xie}
H. Tan, N. Xie, and A. X. Liu, ``An Optimization Framework for Active Physical-Layer Authentication," in IEEE Trans. Mob. Comput., vol. 23, no. 1, pp. 164-179, Jan. 2024. 
   \bibitem{NOMA}
 N. Xie, Q. Zhang, and J. Chen, ``Privacy-Preserving Physical-Layer Authentication for Non-Orthogonal Multiple Access Systems," \emph{IEEE J. Sel. Areas Commun.}, vol. 40, no. 4, pp. 1371-1385, Apr. 2022. 
 \bibitem{encryption_key}
 K. Lin, Z. Ji, Y. Zhang, G. Chen, P. L. Yeoh, and Z. He, ``Secret Key Generation Based on 3D Spatial Angles for UAV Communications," 2021 IEEE Wireless Communications and Networking Conference (WCNC), Nanjing, China, 2021, pp. 1-6. 
\bibitem{encryption_operation}
X. Yan, G. Zhou, Y. Huang, W. Meng, A. T. Nguyen, and H. Huang, ``Secure Estimation Using Partially Homomorphic Encryption for Unmanned Aerial Systems in the Presence of Eavesdroppers," \emph{IEEE Trans. Intell. Veh.}, early access. 
\bibitem{PD1}
H. Yazdani, A. Vosoughi, and X. Gong, ``Achievable Rates of Opportunistic Cognitive Radio Systems Using Reconfigurable Antennas With Imperfect Sensing and Channel Estimation," \emph{IEEE Trans. Cogn. Commun. Netw.}, vol. 7, no. 3, pp. 802-817, Sept. 2021. 
\bibitem{PD2}
W. Hedhly, O. Amin, and M. S. Alouini, ``Benefits of Improper Gaussian Signaling in Interweave Cognitive Radio With Full and Partial CSI," \emph{IEEE Trans. Cogn. Commun. Netw.}, vol. 6, no. 4, pp. 1256-1268, Dec. 2020. 
\bibitem{Shannon_secrecy}
C. E. Shannon, ``Communication Theory of Secrecy Systems", \emph{Bell Syst. Tech. J.}, vol.28-4, page 656--715, Oct. 1949. 
 \bibitem{secrecy_rate}
 M. K. Shukla and H. H. Nguyen, ``Ergodic Secrecy Sum Rate Analysis of a Two-Way Relay NOMA System," \emph{ IEEE Syst. J.}, vol. 15, no. 2, pp. 2222-2225, June 2021. 
\bibitem{DCA}
F. Fang, H. Zhang, J. Cheng, and V. C. M. Leung, ``Energy-Efficient Resource Allocation for Downlink Non-Orthogonal Multiple Access (NOMA) Network,” \emph{IEEE Trans. Commun.}, vol. 64, no. 9, pp. 3722-3732, July 2016. 
\bibitem{AFP}
M. Qaisi, S. Althunibat, and M. Qaraqe, ``Phase-Assisted Dynamic Tag-Embedding Message Authentication for IoT Networks," \emph{ IEEE Internet Things J.}, vol. 9, no. 20, pp. 20620-20629, 15 Oct. 15, 2022.
\bibitem{SIMO}
Z. Gu, H. Chen, P. Xu, Y. Li, and B. Vucetic, ``Physical Layer Authentication for Non-Coherent Massive SIMO-Enabled Industrial IoT Communications," \emph{IEEE Trans. Inf. Forensics Secur.}, vol. 15, pp. 3722-3733, 2020. 
\end{thebibliography}
\end{document}